\documentclass[11pt,a4paper]{article}
\pdfoutput=1
\usepackage{jheppub,tcolorbox}
\usepackage[utf8]{inputenc}
\usepackage{amsmath,amssymb}

\usepackage{tikz} 
\usetikzlibrary{decorations.markings} 
\usetikzlibrary{decorations.text} 
\usetikzlibrary{positioning} 
\usetikzlibrary{backgrounds} 
\usetikzlibrary{fit} 
\usetikzlibrary{calc} 
\usetikzlibrary{shapes} 

\tikzset{dot/.style={draw,circle,inner sep=.7pt,fill,node
    distance=1cm}} 
\tikzset{dot1/.style={draw,circle,inner sep=.7pt,fill}} 
\tikzset{triangle/.style={draw,regular polygon, regular polygon
    sides=3}} 
\tikzset{->-/.style={decoration={
  markings,
  mark=at position .5 with {\arrow{>}}},postaction={decorate}}} 
\tikzset{-<-/.style={decoration={ 
  markings,
  mark=at position .5 with {\arrow{<}}},postaction={decorate}}}

\newcommand\dd{{\rm d}}

\newcommand\cA{{\cal A}}

\newcommand\cV{{\cal V}}

\newcommand\be{\begin{equation}}
\newcommand\ee{\end{equation}}
\newcommand\bea{\begin{eqnarray}}
\newcommand\eea{\end{eqnarray}}
\newcommand\rmd{{\rm d}}

\begin{document}

\begin{titlepage}
\renewcommand{\thefootnote}{\fnsymbol{footnote}}


\vspace*{-2.0cm}

\begin{center}
{\textbf{\huge A String Theory
\vskip0.3cm
Which Isn't About Strings
}}%
\end{center}
\vspace{1.0cm}

\centerline{
\textsc{\large Kanghoon Lee,} $^{a}$%
\footnote{kanghoon.lee1@gmail.com} \hskip0.5cm
\textsc{\large Soo-Jong Rey,} $^{b}$%
\footnote{rey.soojong@gmail.com} \hskip0.5cm  
\textsc{\large J. A.  Rosabal} $^{a}$%
\footnote{j.alejandro.rosabal@gmail.com}
}

\vspace{0.6cm}

\begin{center}
${}^a${\it Fields, Gravity \& Strings @ {\rm CTPU}, Institute for Basic Science, \\
70 Yuseong-daero 1689-gil, Daejeon 34047, \rm KOREA}
\vskip0.2cm
${}^b${\it School of Physics \& Astronomy and Center for Theoretical Physics,\\
Seoul National University, 1 Gwanak-ro, Seoul 08862, \rm KOREA}
\end{center}

\vspace*{1cm}

\centerline{\bf Abstract}

\begin{centerline}
\noindent
Quantization of closed string proceeds with a suitable choice of worldsheet vacuum.  A priori, the vacuum may be chosen independently for left-moving and right-moving sectors. We construct {\sl ab initio} quantized bosonic string theory with left-right asymmetric worldsheet vacuum and explore its consequences and implications. We critically examine the validity of new vacuum and carry out first-quantization using standard operator formalism. Remarkably, the string spectrum consists only of a finite number of degrees of freedom: string gravity (massless spin-two, Kalb-Ramond and dilaton fields) and two  massive spin-two Fierz-Pauli fields. The massive spin-two fields have negative norm, opposite mass-squared, and provides a Lee-Wick type extension of string gravity. We compute two physical observables: tree-level scattering amplitudes and one-loop cosmological constant. Scattering amplitude of four dilatons is shown to be a rational function of kinematic invariants, and in $D=26$ factorizes into contributions of massless spin-two and a pair of massive spin-two fields. The string one loop partition function is shown to perfectly agree  with one loop Feynman diagram of string gravity and two massive spin-two fields. In particular, it does not exhibit modular invariance. We critically compare our construction with recent studies and contrast differences. 
\end{centerline}
\thispagestyle{empty}
\end{titlepage}

\setcounter{footnote}{0}

\tableofcontents

\newpage
\section{Introduction}
String theory, as  a consistent theory of quantum gravity, has grown to its maturity with extensive study over the last five decades. However, due to intricacies involved, it has been difficult to use it to further our understanding of quantum gravity. Nevertheless, it has served a rich source of new theoretical developments.  While the second quantized string field theory needs to be developed further, especially for the closed and supersymmetric cases, one would think that the first quantization of string theory, either in Nambu-Goto \cite{Nambu:1970Copenhagen, Goto:1971ce} or Polyakov \cite{Polyakov:1981rd} formulations, is well developed and thoroughly understood. Still, we would like to go back to the basic starting point and ask ourselves to see if there still is anything to learn more about string theory itself and its premises. In particular, we would like to ask the following questions. 
\begin{itemize}
\item What is string theory? What are the fundamentals of string theory?

\item Can a string theory consist only of a finite number of degrees of freedom?

\item Can one take worldsheet covariance broken at classical or quantum level?

\item By promoting Pauli-Villar regulator to dynamical fields, one obtains the Lee-Wick \cite{Lee:1969fy} extension. Can one construct the Lee-Wick or related alternatives in string theory?

\item The Fierz-Pauli theory \cite{Fierz:1939ix} of massive spin-two fields or multi gravity theory are notoriously difficult. Can they be formulated within (or in terms of) string theory?

\item The double field theory \cite{Siegel:1993th, Siegel:1993xq, Hull:2009mi, Hohm:2010pp, Hohm:2010jy}  is developed for manifest T-duality of string theory when truncated to the massless string gravity (metric, Kalb-Ramond \cite{Kalb:1974yc}, dilaton). Can one construct a string theory whose spectrum just amounts to that of the double field theory?

\item The Gross-Mende \cite{Gross:1987kza}  saddle-point equation for high-energy string scattering and the Cachazo-He-Yuan \cite{Cachazo:2013gna, Cachazo:2013iea, Cachazo:2014nsa} scattering equation for ambitwistor string \cite{Mason:2013sva}  scattering exhibit similar structure. Both originating from string theory, are they identical or merely a coincidence?
\end{itemize}
These questions seem random and unrelated one another. The thesis of this paper is to demonstrate that, to the contrary, the questions above and their answers are all intricately weaved together and the unifying framework is string theory itself!  

So, what is string theory? The relativistic string is an extended object -- elastic and tensile -- so it necessarily includes  infinitely many particle excitations.  That is, if one attempts to describe it in terms of local fields, one must introduce an infinite number of such fields. If one integrates out all but a finite number of fields, one ends up with a field theory but with non-locality at a distance shorter than the string scale. As conventionally formulated,  string theory is quantized with the following properties:
\begin{itemize}
\item
string worldsheet dynamics is invariant under diffeomorphism and Weyl transformations, maintained both at classical and quantum levels. 
\item
string zero-modes and excitations are quantized with the choice of vacuum $|0 \rangle = |0 \rangle_R \otimes |0 \rangle_L$ that is symmetric between the left-moving sector and the right-moving sector,
\begin{tcolorbox}
\leftline{\bf conventional vacuum}
\vskip-0.5cm
\bea
\label{conventionalvac}
&& P_{R}^\mu |0_R \rangle = a^\mu_n \vert 0 _R \rangle = 0 \quad \mbox{and} \quad
\overline{P}_L^{\mu} |0_L \rangle = \overline{a}^\mu_n \vert 0_L \rangle = 0 \qquad (n  = 1, 2, \cdots) \qquad
\eea
viz. the worldsheet vacuum is chosen to be a null element
\bea
\vert 0_R \rangle \in \mbox{Ker} (P_R^{\mu}) \otimes \mbox{Ker} (a^\mu_n) \qquad \mbox{and} \qquad
\vert 0_L \rangle \in \mbox{Ker} (\overline{P}_L^{\mu}) \otimes \mbox{Ker} (\overline{a}^\mu_n), 
\eea
\end{tcolorbox}
and vacuum expectation values of local operators are prescribed with forward time-ordering for both the left-moving and right-moving sectors. 
Note that the vacuum maintains Poincare invariance on the worldsheet (which is the symmetry left after the conformal gauge fixing) and the spacetime, respectively. 
\item 
string spacetime dynamics is ultraviolet finite due to the aforementioned infinitely many spacetime fields and world-sheet modular invariance.
\item
string dynamics, on both worldsheet and spacetime, is unitary; each state has positive norm in the Hilbert space (though the theory is typically afflicted by tachyons in the absence of spacetime supersymmetry).
\end{itemize}
These are features that are not shared by any quantum field theory we are aware of. 

In this work, we challenge this folklore by answering the question "Is it possible to quantize closed string into a theory that does not carry all of the above features?" affirmatively and demonstrate that a quantized string theory can just be a field theory in disguise. Moreover, we will be able to relate the resulting string theory to the seemingly disparate questions raised in the beginning.  We achieve this goal by relaxing several tacit assumptions we make for the conventional string theory quantization. The idea is to adopt the string Fock space vacuum differently from conventional string quantization by choosing the left-moving and right-moving vacua as
\begin{tcolorbox}
\leftline{\bf new vacuum}
\vskip-0.5cm
\bea\label{newvac}
P_R^{\mu} \vert 0_R \rangle = \alpha^\mu_n \vert 0_R \rangle = 0   \quad \mbox{and} \quad \langle 0_L \vert \overline{P}_L^{\mu} = \langle 0_L \vert \overline{\alpha}^\mu_n = 0 \qquad (n = 1, 2, 3, \cdots)
\label{vc1}\eea
viz. the worldsheet vacua are chosen as left null elements
\bea
\vert 0_R \rangle \in \mbox{Ker} (P_R^\mu) \otimes \mbox{Ker} (\alpha^\mu_n) \qquad \mbox{and} \qquad
\langle 0_L \vert \in \mbox{coKer} (\overline{P}_L^\mu) \otimes \mbox{coKer} (\overline{\alpha}^\mu_n).
\eea
\end{tcolorbox}
\noindent
Along with this new choice of the vacuum, we are required to choose time-ordering backward for the left-moving sector, in contrast to forward for the right-moving sector. 

With such a choice of vacuum, we demonstrate that
\begin{itemize}
\item
the theory contains only a finite number of particle excitations, consisting of massless string gravity fields (metric, Kalb-Ramond, dilaton) and a pair of massive Pauli-Fierz spin-two fields of mass-squared $\pm 4/\alpha'$,   
\item the theory is non-unitary; while the massless string gravity sector is unitary, the massive spin-two fields have negative norm 
\footnote{Though the theory may restore the unitarity in the presence of spacetime supersymmetry \cite{Leite:2016fno}-\cite{Huang:2016bdd}.},
\item the new vacuum respects spacetime Poincare invariance but spontaneously breaks worldsheet Poincare invariance. Rotational symmetry of Euclidean worldsheet is broken. Accordingly, worldsheet modular transformation is no longer a symmetry.

\end{itemize}
\begin{figure}[hbt]
\centering
  \includegraphics[width=1\textwidth]{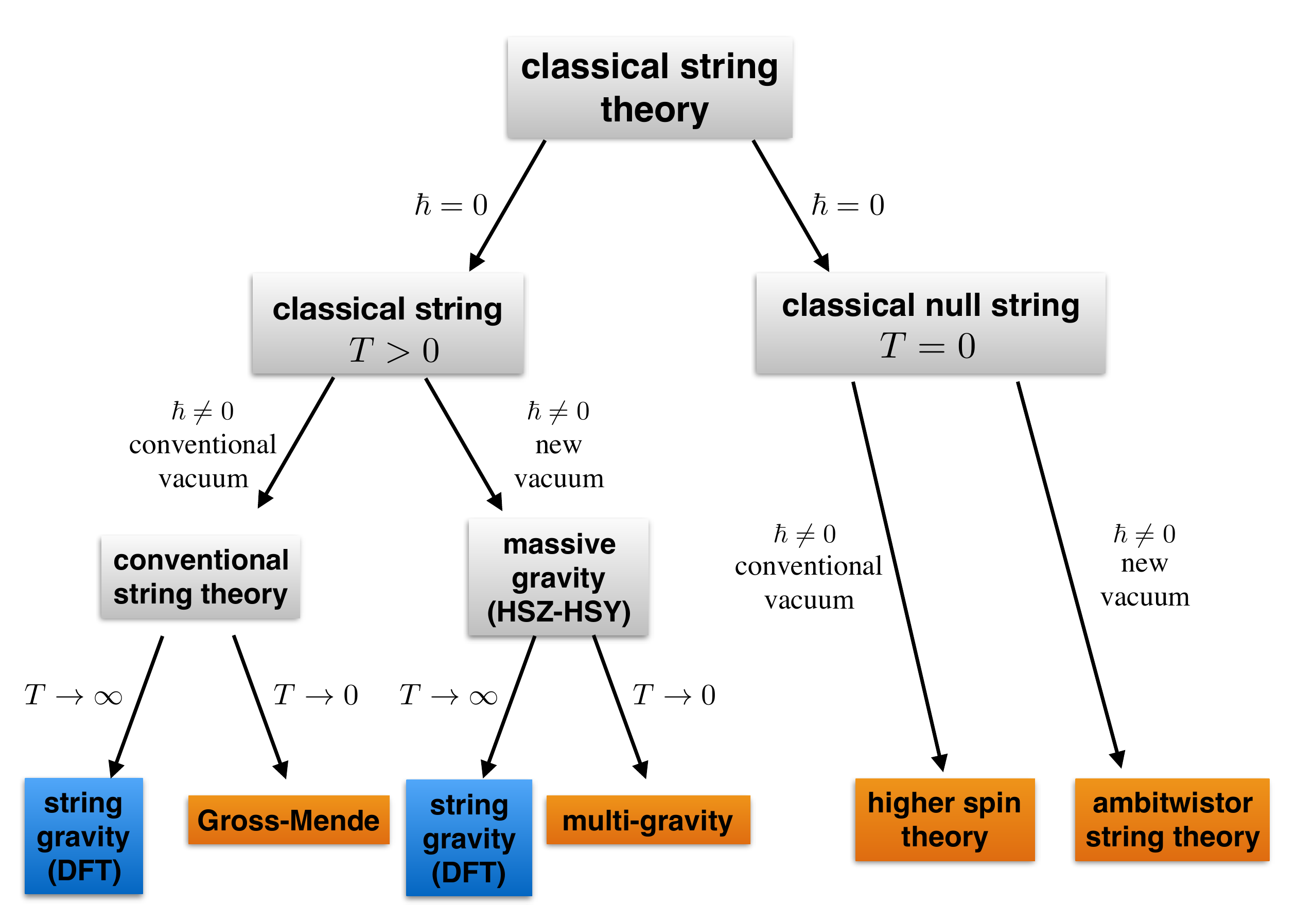}
  \caption{\sl Various limits of quantized closed string theory. Different limits correspond to different order of the first-quantized worldsheet $\hbar$ and the string tension $T = (2 \pi \alpha')^{-1}$ (relative to a characteristic energy scale).  It should be emphasized that $\hbar$ and $T$ limits do not commute. Ambitwistor string differs from multi-gravity, and a precise relation between higher spin theory and Gross-Mende high-energy scattering remains unsettled. }
\end{figure}
\hfill\break

Like the quantization over the conventional vacuum, this quantization still contains two parameters: the worldsheet Planck constant $\hbar$ and the string tension $T = 1/(2 \pi \alpha')$ \footnote{We emphasize that these two parameters should be distinguished and differentiated. A classical string can and do have a finite tension, and a tensionless string can and should be first-quantized.}. By taking various regimes of these two parameters for the closed bosonic string quantized over either conventional or new vacuum, we are able to construct a variety of further simplified theories. We illustrate them in Figure 1. 
Here, we list various quantum field theories indicated in Figure 1 and how they are related to the quantized string theory in either choice of the vacuum.
\begin{itemize}
\item {\bf string gravity (double field theory)}: This theory comprises of the metric, Kalb-Ramond, and dilaton fields. As is well-known, we can obtain this theory by first-quantizing the tensile bosonic string over the conventional vacuum (\ref{conventionalvac}) and then taking the infinite tension limit $T \rightarrow \infty$ at a finite characteristic energy scale.  
\item {\bf Gross-Mende string}: In this regime, first studied by Gross and Mende \cite{Gross:1987kza}, the first-quantized string over the conventional vacuum (\ref{conventionalvac}) is taken to infinite characteristic energy scale relative to the string tension. Alternatively, this is the regime where the string tension vanishes, $T \rightarrow 0$ at a finite characteristic energy scale.
\item {\bf massive gravity}: If a tensile string is quantized over the new vacuum (\ref{newvac}) and keep the tension finite, this theory contains only a finite number of states: the string gravity and a pair of massive Pauli-Fierz spin-two ghost fields. This theory is closely related to the $\alpha'$-corrected double field theory developed by Hohm, Siegel and Zwiebach (HSZ) \cite{Hohm:2013jaa} and to the modified Kawai-Lewellin-Tye (KLT) relation  \cite{Kawai:1985xq,Leite:2016fno,Li:2017emw} developed by Huang, Siegel and Yuan (HSY)  \cite{Huang:2016bdd} 
\footnote{This development is built upon earlier worldsheet approach in \cite{Siegel:2015axg}.}.  
\item {\bf multi-gravity}: If the tension in the massive gravity is taken to zero $T \rightarrow 0$ at a finite characteristic energy scale,  the resulting theory is a fully interacting multi-gravity theory of  string gravity and a pair of massless spin-two ghost fields. 
\item {\bf higher spin theory}: If the tension is first taken to zero $T \rightarrow 0$ and then quantize the tensionless or null string over the conventional vacuum (\ref{conventionalvac}), the resulting theory is the higher-spin gauge theory of infinitely many massless fields of spin two and higher \cite{Karlhede:1986wb,Lindstrom:1990qb,Lindstrom:1990ar,Isberg:1993av} .  
\item {\bf ambitwistor theory}: If the tension is first taken to zero $T \rightarrow 0$ and then quantize the tensionless or null string over the new vacuum (\ref{newvac}), the resulting theory is the ambitwistor theory \cite{Mason:2013sva} which was developed to explain a string theory origin of the scattering equation proposed by Cachazo, He and Yuan (CHY) \cite{Cachazo:2013gna}-\cite{Cachazo:2014nsa}. 
\end{itemize}
So, one may regard the quantized bosonic string we develop in this paper as {\sl ab initio} formulation of all these limiting field theories. 

We should mention that the possibility of different worldsheet vacuum choice was considered by Hwang, Marnelius and Saltsidis \cite{Hwang:1998gs} in the quantization of tensile string. However, they dismissed the new vacuum (\ref{newvac}) on the basis that the new vacuum is not a physical state. This is in agreement with the result we find in this paper by a different method: there is no scalar tachyon in the spectrum. They also argued that massless string gravity states are negative norm ghosts. Here, we differ from theirs. We find that the new vacuum itself, which is not a physical state,  necessarily has a negative norm and in turn the massless string gravity states have positive norm. There are also a pair of massive spin-two states of negative norm and opposite mass-squared, which act as a variant of the Lee-Wick ghost fields. So, our viewpoint differs from them in that this issue is far more general and deserves the merit of its own. What we find is that, by reassessing possible choices of string worldsheet vacuum and quantizing string in all the vacua,  we obtain quantum theories whose degrees of freedom are drastically reduced from the conventional string theory, viz. from infinite to all disappeared but a few. 

The different worldsheet vacua were also considered previously by Gamboa, Ramirez and Ruiz-Altaba \cite{Gamboa:1989px,Gamboa:1989zc} in the quantization of tensionless or null string. The choice of new vacuum was recently revisited in the work by Casali and Tourkine \cite{Casali:2016atr}, which nicely clarified the origin of CHY scattering equation via the ambitwistor string. Originally, ambitwistor string model was interpreted as an infinite tension limit $T \rightarrow \infty$  of the conventional string theory. On the other hand, the scattering equations in CHY formula was considered to be the saddle point equations in the Gross-Mende limit, which requires the tensionless limit. Casali and Tourkine resolved this discrepancy by showing  that instead of the infinite tension limit,  ambitwistor string should be considered as tensionless or null string quantized over the new vacuum (\ref{newvac}), which they renamed as ambitwistor vacuum. 

We organize this paper as follows. In section 2, we recapitulate the first quantization of closed bosonic string over the conventional vacuum.  We also recall two recent attempts to modify string theory and discussing their potential pitfalls. In section 3, we undertake the first quantization of closed bosonic string over the new vacuum (\ref{newvac}). We first identify the correct time-ordering on the worldsheet: the ordering is such that the right-moving sector takes forward time-ordering (as in the conventional vacuum) while the left-moving sector takes backward time-ordering. and from it compute the two-point correlation function. We also check that the anti-time-ordering is compatible with the normal ordering in the tensionless limit  \cite{Casali:2016atr,Gamboa:1989px,Gamboa:1989zc}.  We further analyze the representation theory of Virasoro algebra over the new vacuum, and demonstrate that the spectrum consists only of string gravity (metric, Kalb-Ramond, dilaton) and a pair of massive spin-two ghost fields. We also compute Regge intercept, critical dimension and central charge of the left-moving and right-moving sectors, first from heuristic argument and then from careful treatment of the BRST quantization of $b, c$ ghosts.  

In section \ref{amplitudes}, we study string interactions. We first present the generating function for computing the tree-level scattering amplitude, which is computed within the operator formalism. We then compute the four-point dilaton scattering amplitude. As expected for a field theory with finite field contents, the string amplitude is a rational function of finite order polynomials of Mandelstam invariants. We study factorization property of the amplitude and find perfect agreement between the pole structure of the amplitude and the mass spectrum in section \ref{spectrum} provided the spacetime dimension is set to 26. 
In section \ref{partition_func}, we compute the one-loop vacuum amplitude, viz. string partition function on torus worldsheet. We confirm that the partition function properly counts the degrees of freedom in the theory, but it is not modular invariant. We attribute the lack of modular invariance to the fact that the new vacuum spontaneously breaks the worldsheet Poincare invariance, which in turn leads to states with negative norm for the left-moving sector. Conclusions and outlooks are presented in section \ref{conclusions}.

In Appendix A, we study the tensionless limit for a quantized string over the new vacuum. We recall that the tensionless limit is defined by the rescaling of worldsheet time $\tau \rightarrow \epsilon \tau$ and of string tension $T \rightarrow \epsilon T_0$ for a fiducial, nonzero tension $T_0$ and then sending $\epsilon \rightarrow 0$. Taking this limit, we show that we can reconstruct mode expansion of tensionless or null string, that the Virasoro algebra over the new vacuum is reduced to the Galilean conformal algebra, and that the spectrum in this limit matches with that of tensionless or null string. In Appendix B, we collect the three-point amplitudes between two dilatons and a string gravity field or between two dilatons and a massive spin-2 field, $a_{\mu \nu}$ or $\overline a_{\mu \nu}$. Combining the three-point amplitudes, we reconstruct four-point amplitudes at each factorization channel. We find that the result perfectly agrees with the four-point amplitude we computed in section \ref{amplitudes} directly from string theory.
Details of the four-point dilaton amplitude are summarized  in Appendix C.
 

\section{Quantization of Closed String}
The core of this paper is to challenge the conventional route to the quantized string theory. So, we shall begin our considerations from the basics of string theory. In this section, we redo the first-quantization of closed bosonic string, paying special attention to the choice of worldsheet vacua for left-moving and right-moving sectors as well as center-of-mass zero modes. 

\subsection{Conventional route: quantization over conventional vacuum}\label{Conventional string quantization}

Our starting point is the Polyakov formulation of closed bosonic string theory \cite{Polyakov:1981rd}, whose worldsheet action is given by
\bea
S = - {1 \over 4 \pi \alpha'} \int_\Sigma \rmd \tau \rmd \sigma \, \sqrt{-h} h^{ab} \partial_a X^\mu \partial_b X^\nu \eta_{\mu \nu} \ .\label{action0}
\eea
Here, $\Sigma$ is the Lorentzian worldsheet parametrized by $(\tau, \sigma)$. The worldsheet action is a functional of the metric $h_{ab}$ and the scalars $X^\mu$.
This action is invariant under the worldsheet  diffeomorphism. This local symmetry is fixed by imposing the conformal gauge condition, $\sqrt{-h} h^{ab} = \eta^{ab} = \mbox{diag}(-1, 1)$. The equations of motion for the worldsheet metric is then reduced to the constraints
\bea
T_{ab} = \partial_{a} X^{\mu} \partial_{b} X^{\nu} - \tfrac{1}{2} \eta_{ab} \eta^{cd}\partial_{c} X^{\mu} \partial_{d}X^{\nu}  = 0\,.\label{viracons}
\eea
Later, we will separately treat the Faddeev-Popov ghosts and the BRST quantization. The canonical momenta $\Pi^\mu$ conjugate to $X^{\mu}$ are given by
\begin{equation}
  \Pi^{\mu}(\tau,\sigma) = \frac{1}{2\pi \alpha'} \partial_{\tau}X^{\mu}(\tau,\sigma)\,.
\label{}\end{equation}

In the conformal gauge, the remaining fields on the worldsheet are the string coordinates $X^\mu(\sigma, \tau)$. Their equations of motion in the conformal gauge read 
\bea
\Box X^\mu (\tau, \sigma) = \left( \partial_\tau^2 - \partial_\sigma^2 \right) X^\mu(\tau, \sigma) = 0\, \qquad (\mu = 0, 1, \cdots, D-1).
\eea
We impose the periodic boundary condition in the $\sigma$ direction,
\bea
X^\mu (\tau, \sigma) = X^\mu (\tau, \sigma + 2 \pi), 
\qquad
\Pi^\mu (\tau, \sigma) = \Pi^\mu (\tau, \sigma + 2 \pi) \, , 
\eea
and find the most general closed string solution as a sum of arbitrary left-moving  and right-moving profiles
\bea
X^\mu (\tau, \sigma)= X_L^\mu (\tau + \sigma) + X_R^\mu (\tau - \sigma).
\eea
Each of them are not necessarily periodic in $\sigma$ but their sum should be. Expanding the two functions into zero mode and harmonic modes,  
\bea
X_L^\mu(\tau, \sigma) &=& 
 \frac{1}{2}X_0^\mu + {\alpha' \over 2} P^\mu (\tau + \sigma) + \sqrt{\alpha' \over 2} \sum_{n \ne 0} {i \over n} \overline \alpha^\mu_n  e^{ -  i n (\tau + \sigma)}\,,
\\
X_R^\mu (\tau, \sigma) &=&
 \frac{1}{2}X_0^\mu + {\alpha' \over 2} P^\mu (\tau - \sigma) + \sqrt{\alpha' \over 2}  \sum_{n \ne 0} {i \over n} \alpha^\mu_n  e^{ -  i n (\tau - \sigma)}\ . 
\eea
The zero-mode part describes rigid motion of closed string,
\bea
{1 \over 2} X_0^\mu + {\alpha' \over 2} P^\mu (\tau + \sigma) + {1 \over 2} X_0^\mu + {\alpha' \over 2} P^\mu (\tau - \sigma)
 = X_0^\mu + \alpha' P^\mu \, \tau,
\eea
and trivially periodic, as it should be. 
The canonical momentum $\Pi^\mu$ can also be decomposed to left-moving and right-moving sectors, $\Pi^\mu = \Pi^\mu_L + \Pi^\mu_R$, as
\bea\label{canonicalmomentum}
\Pi_L^\mu (\tau, \sigma) &=& {1 \over 2 \pi} \left[ {1 \over 2} P^\mu  + \sqrt{ 1 \over 2 \alpha'} \sum_{n \ne 0} \overline \alpha_n^\mu e^{-in (\tau + \sigma)} \right] \,,
\nonumber \\
\Pi_R^\mu (\tau, \sigma) &=& {1 \over 2 \pi} \left[{1 \over 2} P^\mu + \sqrt{1 \over 2 \alpha'} \sum_{n \ne 0} \alpha_n^\mu e^{-in(\tau  - \sigma)} \right] \ .
\eea

The Lorentzian worldsheet can be Wick-rotated to an Euclidean plane by a conformal mapping, and then Wick-rotated back to a Lorentzian cylinder 
\bea
z = \exp i(\tau - \sigma), \qquad \bar z = \exp i(\tau + \sigma).\label{ztrans}
\eea
In terms of $z, \overline z$, the mode expansions are given by
\bea\label{modeex}
X_L^\mu (\bar z) &=& X_{0L}^\mu - i {\alpha' \over 2} P_L^\mu \log \bar z + \sqrt{ \alpha' \over 2} \sum_{n \ne 0} {i \over n} \overline \alpha_n^\mu \bar z^{-n}\,,
\nonumber \\
X_R^\mu (z) &=& X_{0R}^\mu - i {\alpha' \over 2} P_R^\mu \log z + \sqrt{\alpha' \over 2} \sum_{n \ne 0} {i \over n} \alpha_n^\mu z^{-n}\ ,
\eea
viz. a pair of left-moving and right-moving chiral bosons. Here, keeping in mind of the situation that some of the spacetime directions are compactified and of the double field theory formulation therein, we are considering the most general case where $X^\mu_{0L}$, $X^\mu_{0R}$, $P^\mu_L$ and $P^\mu_R$ are independent zero modes.  If we restrict our attention to  $X^\mu_{0L}=X^\mu_{0R}$,   $P^\mu_L=P^\mu_R$ and vertex operators are constructed only from the sum $(X^\mu_L+X^\mu_R)$, the dynamics would be reduced to  string theory in a noncompact  spacetime.

We now quantize the world-sheet dynamics. Upon quantization, $X_0^\mu, P^\mu, \alpha^\mu_n, \overline\alpha^\mu_n$ are promoted to operators. Accordingly, equations of motion and Virasoro constraints are promoted to operator equations.  We proceed with the canonical quantization formalism by promoting classical Poisson bracket of conjugate variables $(X^\mu(\tau, \cdot), P^\mu (\tau, \cdot))$ to quantum commutation relations
\be
[X^{\mu}(\tau,\sigma), \Pi^{\nu}(\tau,\sigma^{\prime})]=i\ \eta^{\mu\nu}\delta(\sigma-\sigma^{\prime})\label{canonicaldonttouchit}\,. 
\ee
Mode expanding according to (\ref{modeex}), the zero-mode obeys the commutation relations
\be
[X_0^\mu, P^\nu] = i \eta^{\mu \nu} ,\label{cr1}
\ee
while the harmonic modes obey
\begin{equation}
\begin{aligned}
  &[\overline \alpha^\mu_m, \overline \alpha^\nu_{n}] = m\, \delta_{m+n,0}\, \eta^{\mu \nu}\,,
  \\
  &[ \alpha^\mu_m, \overline \alpha^\nu_{n}] = 0\,,
  \\
  &[\alpha^\mu_m, \alpha^\nu_{n}] =m \, \delta_{m+n,0} \, \eta^{\mu \nu} \,.
\end{aligned}\label{cr2}	
\end{equation}
Mode expanded, the Virasoro constraints also give rise to an infinite set of operator conditions. For zero mode, the operators are
\bea
L_0 = \left( {1 \over 2} \alpha' p^2 
+ \sum_{n=1}^\infty \alpha_{-n} \cdot \alpha_n \right) - a\,,
\nonumber \\
\overline L_0 = \left( {1 \over 2} \alpha' p^2  
+ \sum_{n=1}^\infty \overline \alpha_{-n} \cdot \overline \alpha_n \right) - \overline a\,.
\label{uViraZero}\eea
Here, $a$ and $\overline{a}$ are so-called intercept constants that are to be fixed from quantum consistency.  For non-zero modes, the Virasoro operators are
\begin{equation}
  L_m = \frac{1}{2}\sum_{n=-\infty}^{\infty}\alpha_{m-n}\cdot\alpha_n \quad
  \mbox{and} \quad 
\overline{L}_m = \frac{1}{2}\sum_{n=-\infty}^{\infty}\overline{\alpha}_{m-n}\cdot\overline{\alpha}_n\, \quad \text{for}~m\neq 0\,.
\label{}\end{equation}

So far, all equations are operator-valued, and so they do not depend on the choice of the world-sheet vacuum. We now choose a vacuum of the quantum string %
\bea
\vert 0 \rangle = \vert 0 \rangle_0 \otimes \vert 0 \rangle_L \otimes \vert 0 \rangle_R\ 
\eea
and construct the Fock space of excited string states by acting creation operators on the vacuum state. Conventionally one chooses the vacuum according to
\begin{equation}
  P^{\mu} |0\rangle_{0} = 0\, ,
\label{}\end{equation}
for the zero mode, and
\bea
\alpha_n^\mu | 0 \rangle_{R} = 0\,,\quad \quad  \overline{\alpha}_n^\mu | 0 \rangle_{L} = 0\,,\qquad \text{for}~n>0 \label{con_vac}
\eea
for harmonic modes. This choice of vacuum treats the excitations symmetrically between the left-moving sector and the right-moving sector. Furthermore, the time ordering is taken forward, putting operators in the past to the left and operators in the future to the right. The Fock space constructed out of these choices of vacuum and time-ordering is infinite-dimensional. Moreover, in this conventional vacuum, the intercept constants $a, \bar a$ are determined to be 1, rendering the string gravity states massless. As is well known, the Fock space states form an infinite tower of string excitations, forming the Regge spectrum. We also recall the Virasoro algebra of left-moving and right-moving sectors
\bea \nonumber\label{convvirasoroalgebra}
\big[L_m, L_n\big ] & = & (m-n)L_{m+n}+\frac{c}{12}m(m^2-1)\delta_{m+n,0}, \\
\big[\overline{L}_m, \overline{L}_n\big ] & = & (m-n)\overline{L}_{m+n}+\frac{\bar c}{12}m(m^2-1)\delta_{m+n,0} \, . 
\eea
where the central charges $c, \bar c$ take equal values, $c = \bar c$, as a consequence of the vacuum choice (\ref{con_vac}). 

All being well, we next would like to consider different choice of the worldsheet vacuum.  To motivate such choices, we first recapitulate two important relevant works that take us to unconventional routes and discuss them in the context of quantized tensile string theory. 


\subsection{Unconventional route: metric sign flip prescription}\label{Unconventional string quantization1}
Recently, within the KLT framework of building closed string theory out of double copies of open string, Huang, Siegel and Yuan \cite{Huang:2016bdd} proposed to treat the commutation relations of left-moving and right-moving sectors oppositely,
\be
[X_{0L}^\mu, P_L^\nu] = -i \eta^{\mu \nu}, \qquad \qquad [X_{0R}^\mu, P_R^\nu] = i \eta^{\mu \nu} ,\label{wcr1}
\ee
\be
 [\overline \alpha^\mu_m, \overline \alpha^\nu_{n}] = -m\delta_{m+n,0}\eta^{\mu \nu} , 
\quad
[\alpha^\mu_m, \alpha^\nu_{n}] =m \delta_{m+n,0} \eta^{\mu \nu} .\label{wcr2}
\ee
They choose the conventional vacuum, viz. $P^\mu_L\vert 0 \rangle=P^\mu_R\vert 0 \rangle=0$ and $\overline \alpha^\mu_n\vert 0 \rangle=\alpha^\mu_n\vert 0 \rangle=0$ for $n>0$.

To see if Eq.(\ref{wcr2}) is compatible with quantized closed string, we 
explore consequences of the HSY prescription on the conjugate pair of closed string $X^{\mu}(\sigma, \tau)$ and $\Pi^{\mu}(\tau, \sigma)$  and check the compatibility with the canonical commutation relations (\ref{canonicaldonttouchit}). We thus return to the mode expansion in the most general form in (\ref{modeex}), 
\bea\label{t_mode_exp}
X_L^\mu (\bar z) &=& X_{0L}^\mu - i {\alpha' \over 2} P_L^\mu \log \bar z + \sqrt{ \alpha' \over 2} \sum_{n \ne 0} {i \over n} \overline \alpha_n^\mu \bar z^{-n}\,,
\nonumber \\
X_R^\mu (z) &=& X_{0R}^\mu - i {\alpha' \over 2} P_R^\mu \log z + \sqrt{\alpha' \over 2} \sum_{n \ne 0} {i \over n} \alpha_n^\mu z^{-n}\ .
\eea
Plugging the mode expansion (\ref{t_mode_exp}) in the commutation relations for the left-moving and right-moving sectors separately and taking into account (\ref{wcr1}) and (\ref{wcr2}),  we get
\be
[X_R^{\mu}(\tau,\sigma), \Pi_R^{\nu}(\tau,\sigma^{\prime})]= + \frac{1}{2}i\ \eta^{\mu\nu}\delta(\sigma-\sigma^{\prime})\ ,\label{ccr1}
\ee
\be
[X_L^{\mu}(\tau,\sigma), \Pi_L^{\nu}(\tau,\sigma^{\prime})]=-\frac{1}{2}i\ \eta^{\mu\nu}\delta(\sigma-\sigma^{\prime})\ . \label{ccr2}
\ee
This is just the statement of HSY prescription, flipping the spacetime metric sign between the right-moving sector and left-moving sector. However, starting from (\ref{ccr1}) and (\ref{ccr2}), one finds it impossible to obtain the canonical commutation relations (\ref{canonicaldonttouchit}) of the closed string \footnote{From the defining relations $X^{\mu}=X^{\mu}_L+X^{\mu}_R$ and $\Pi^{\mu}=\Pi^{\mu}_L+\Pi^{\mu}_R$, one instead finds
\be
[X^{\mu}(\tau,\sigma), \Pi^{\nu}(\tau,\sigma^{\prime})]=[X^{\mu}(\tau,\sigma)_L+X^{\mu}(\tau,\sigma)_R\ , \ \Pi^{\mu}_L(\tau,\sigma^{\prime})+\Pi^{\mu}_R(\tau,\sigma^{\prime})]=0\label{class1}\ .
\ee
}\ .


\subsection{Unconventional route: tensionless or null string}\label{Unconventional string quantization2}
Another unconventional route is to quantize tensionless or null string. 
To this end, one finds it convenient to express the harmonic mode oscillators in terms of conjugate pairs of Hermitian operators $X_n^\mu$ and $\Pi^\mu_n$,
\begin{equation}
\begin{aligned}
  \alpha_n^{\mu} & = \frac{1}{2\sqrt{T}} \Pi_{n}^{\mu}-in\sqrt{T}X_n^{\mu}\, , 
\\
\overline{\alpha}_n^{\mu} & = \frac{1}{2\sqrt{T}}\Pi_{-n}^{\mu}-in\sqrt{T}X_{-n}^{\mu}\,, 
\end{aligned}\label{C_A_op}
\end{equation}
where $T = 1/(2 \pi \alpha')$ is the string tension.

When quantizing tensionless or null string, as first pointed out in \cite{Gamboa:1989px,Gamboa:1989zc}, there are two possible choices of the vacuum. The first one is the so-called higher spin vacuum. It turns out that this vacuum just descends from the tensionless limit $T \rightarrow 0$  of the conventional vacuum in quantized tensile string. Indeed, by taking $T\to0$ limit in (\ref{C_A_op}), the conventional vacuum (\ref{con_vac}) is reduced to 
\begin{equation}
  \Pi^{\mu}_{n}|0\rangle = 0\,, \qquad \text{for }~n \in \mathbb{Z}\,.
\label{}\end{equation}
In this higher spin vacuum, the mass spectrum is continuous from the outset, and as such there is no critical dimension. 

The second choice of tensionless or null string vacuum is realized by the unconventional vacuum
\begin{equation}
\begin{aligned}
  \Pi_n^{\mu}|0\rangle  = 0\,, \qquad X_n^{\mu}|0\rangle = 0\,, \qquad\quad \text{for}~ n>0\,,
\end{aligned}\label{wrongvac0}
\end{equation}
together with the normal ordering prescription that puts all $X_{n}^\mu$ and $\Pi_{n}^\mu$  with $n > 0$ to the right. This vacuum exhibit several intriguing properties. The spectrum of quantized tensionless or null string consists of a finite number of degrees of freedom and they are all massless.  Moreover, the critical dimension is exactly the same as for the tensile string theory, viz. 26 for bosonic string and 10 for superstring.
Recently, Casali and Tourkine \cite{Casali:2016atr} revisited the quantized tensionless or null string, and argued that the ambitwistor string developed for string theoretic understanding of CHY scattering equation is nothing but the tensionless or null string quantized over the unconventional vacuum (\ref{wrongvac0}). 

In the quantization of tensile string, assuming that the tensionless limit is smooth and analytic, the unconventional vacuum (\ref{wrongvac0}) can be translated to the conditions 
\bea
\alpha_n \vert 0 \rangle_R = 0, \qquad \overline \alpha_{-n} \vert 0 \rangle_L = 0 \qquad \text{for}~~n > 0\label{wrongvac},
\eea
and the normal ordering prescription puts all $\alpha_{n}^\mu$ and $\overline \alpha_{-n}^\mu$ with $n > 0$ to the right, for example,
\begin{equation}
\mathclose:\alpha_{m} \alpha_{-n} \alpha_{p}\mathclose: = \alpha_{-n} \alpha_{m} \alpha_{p} \quad \mbox{and} \quad \mathclose:\overline{\alpha}_{m}\overline{\alpha}_{-n}\overline{\alpha}_{p}\mathclose: = \overline{\alpha}_{m} \overline{\alpha}_{p} \overline{\alpha}_{-n}\,, \qquad \mbox{for} ~ m,n,p>0\,.
\label{norm_ordering}\end{equation}
Note that  Eq.(\ref{wrongvac0}) and Eq.(\ref{wrongvac}) are equivalent in the tensionless limit; in the tensile regime, however, Eq.(\ref{wrongvac0}) and Eq.(\ref{wrongvac}) are not equivalent. 

The discussion so far encompasses what \cite{Gamboa:1989px,Gamboa:1989zc} originally studied in the context of tensionless and null string and then \cite{Casali:2016atr} further studied in the context of ambitwistor string. Both has left the vacuum state for $n=0$ unspecified. We emphasize that some extra care is imperative for the zero-mode sector in order to fully specify the quantized string. In the next section, we will consider the choice of zero-mode vacuum  in great detail and find surprising new information elucidating internal consistency of string quantization. 

Now, one can also find a relation between the unconventional HSY prescription (\ref{wcr1}, \ref{wcr2}) and the unconventional vacuum choice (\ref{wrongvac}) when they are extended to tensile string. One may rename the oscillators as $\overline{\alpha}_{-n}\rightarrow \overline{a}_{n}$ and $\overline{\alpha}_{n}\rightarrow \overline{a}_{-n}$. This way, the vacuum gets defined in the conventional way (\ref{con_vac}) but the commutation relation for the left-moving sector flips the sign, viz.
\be
\big[\overline{\alpha}_m^{\mu},\overline{\alpha}_{-m}^{\nu}\big]  =  m\eta^{\mu\nu}\qquad \rightarrow \qquad \big[\overline{a}_m^{\mu},\overline{a}_{-m}^{\nu}\big]  =  -m\eta^{\mu\nu}\ .
\ee
Clearly, this transformation is not a Bogoliubov transformation, as it changes the commutation relation. However, by performing this non-unitary transformation, we can always relate the unconventional vacuum to the HSY prescription.


\section{Quantized String over the New Vacuum}\label{Sec:QUVC}
In this section, we quantize a closed bosonic string over the new vacuum whose harmonic modes were already deduced in Eq.(\ref{wrongvac}). As emphasized there, the new vacuum is completely specified only after the vacuum of zero-mode is also prescribed. In this section, we will isolate the specification and identify the new vacuum to be 
\bea
 P^\mu_{R}|0\rangle_{0_{R}} = 0\,, && \qquad \sideset{_{0_{L}\!\!}}{}{\mathop{\langle}}0| P^\mu_{L} = 0\, , \nonumber \\
\alpha_n \vert 0 \rangle_R = 0, && \qquad \overline \alpha_{-n} \vert 0 \rangle_L = 0 \qquad \text{for}~~n > 0\label{wrongvacuum}.
\eea
We do this as follows. The equal-time commutation relations are defining operator relations, so they are the same as the conventional string theory. On the other hand, time ordering, as probed by vacuum expectation values of time-evolved operators, would depend on the choice of vacuum. Here, by requiring that two-point correlator of string coordinates is well-defined, we extract a consistent time-ordering prescription for the new vacuum and, from the prescription, identify the correct zero-mode vacuum. We also formulate quantization of the $b$, $c$ ghosts over the new vacuum and determine the string intercept constants.


\subsection{Problem with conventional time-ordering}
Our first step is to construct the two-point correlation $G(i, j)$ of the string coordinates $X^\mu(z, \bar z)$, 
\begin{equation}
G(i, j) \equiv \frac{1}{\left\langle 0|0\right\rangle} {\left\langle 0|\mathclose X^{\mu}(z_{i},\bar{z}_{i}) X^{\nu}(z_{j},\bar{z}_{j})\mathclose |0\right\rangle} \ , 
\label{2pointcorrelator}\end{equation}
over the new vacuum in a well-defined manner. From the mode expansion (\ref{modeex}) and the new vacuum  (\ref{wrongvac}), one finds that the two-point correlation is decomposed into three parts {\sl viz.} contributions of right-moving oscillators, left-moving oscillator and zero-mode sectors,
\begin{equation}
\begin{aligned}
    &G(i, j) 
   = \frac{1}{\sideset{_{R}}{_{R}}{\mathop{\left\langle 0|0\right\rangle}}} {\sideset{_{R}}{_{R}}{\mathop{\left\langle{0}|\mathclose X{}^{\mu}_{R}(z_{i}) X{}_{R}^{\nu}(z_{j})\mathclose  |0\right\rangle}}} + \frac{1}{\sideset{_{L}}{_{L}}{\mathop{\left\langle 0|0\right\rangle}}}	
 {\sideset{_{L}}{_{L}}{\mathop{\left\langle{0}|\mathclose X{}_{L}^{\mu}(\bar{z}_{i}) X{}_{L}^{\nu}(\bar{z}_{j})\mathclose  |{0}\right\rangle}} }    \\
    & \qquad\quad+ \frac{1}{\sideset{_{0}}{_{0}}{\mathop{\left\langle 0|0\right\rangle}}} {{{}_{}}_{}}_{0}\big\langle{0}| \Big(X^{\mu}_{0}-i\frac{\alpha'}{2} P^{\mu}\log\big({z}_{i} \bar{z}_{i}\big)\Big) \Big(X^{\mu}_{0}-i\frac{\alpha'}{2} P^{\mu} \log\big({z}_{j} \bar{z}_{j}\big)\Big)|{0}\big \rangle_{0} \ , 
\end{aligned}\label{correlator1}
\end{equation}
where $X{}^{\mu}_{R}$ and $X{}^{\mu}_{L}$ are the left-moving and right-moving parts of nonzero modes of string coordinate $X^\mu$ ,
\begin{equation}
  X^{\mu}_{R} (z) =  i \sqrt{\alpha' \over 2} \sum_{n \ne 0} {1 \over n} \alpha^\mu_n z^{-n}\,, \qquad 
  X^{\mu}_{L} (\overline{z}) = i \sqrt{\alpha' \over 2} \sum_{n \ne 0} {1 \over n} \overline{\alpha}^\mu_n\, \bar{z}^{-n}.
\label{}\end{equation}
To further proceed, we take the worldsheet to be the Euclidean plane, obtained from the Lorentzian cylinder by the Wick rotation $\tau \rightarrow i \tau$. We denote the coordinates of Euclidean plane as $z, \bar z$. 
 
The contribution of right-moving oscillators is the same as that in the conventional quantization. Substituting the above mode expansion of $X^{\mu}_{R}(z)$ in (\ref{modeex}), we have
\begin{equation}
\begin{aligned}
  \sideset{_{R}}{_{R}}{\mathop{\left\langle{0}|\mathclose X^{ \mu}_{R}(z_{i}) X_{R}^{ \nu}(z_{j})\mathclose |{0}\right\rangle}} 
  = {{{}_{}}_{}}_{R}\big\langle{0}| \Big(i\sqrt{\alpha' \over 2}\sum_{n>0}\frac{1}{n} \alpha^{\mu}_{n} z_{i}^{-n}\Big) \Big(i\sqrt{\alpha' \over 2}\sum_{n>0}\frac{1}{-m} \alpha^{\nu}_{-m} z_{j}^{m}\Big)|{0}\big \rangle_{R}\,. 
\end{aligned}\label{}
\end{equation}
So, the contribution of right-moving sector is reduced to
\begin{equation}
\begin{aligned}
    \frac{\alpha'}{2}\sum_{m,n>0}\frac{1}{mn} \frac{z^{m}_{j}}{z^{n}_{i}} \sideset{_{R}}{_{R}}{\mathop{\left\langle 0| \alpha^{\mu}_{n}\alpha^{\nu}_{-m}|0\right\rangle}}
  &= + \eta^{\mu\nu}\frac{\alpha'}{2} \sum_{n>0} \frac{1}{n}\Big(\frac{z_j}{z_i}\Big)^{n}\sideset{_{R}}{_{R}}{\mathop{\left\langle 0|0\right\rangle}} 
  \\
  &= -\eta^{\mu\nu}\frac{\alpha'}{2} \log\Big(1-\frac{z_j}{z_i}\Big)\sideset{_{R}}{_{R}}{\mathop{\left\langle 0|0\right\rangle}}\,. 
\end{aligned}\label{right_correlator}
\end{equation}
Defined on the Euclidean worldsheet, the series is convergent and the resummation is well-defined provided the ordering is taken forwardly as $|z_{i}|>|z_{j}|$. 
This convergence condition is consistent with the conformal time ordering, so it implies that the normal ordering and the forward time ordering are equivalent.

The contribution of left-moving oscillators requires a careful treatment. Substituting the above mode expansion for $X^{\mu}_{L}(\overline{z})$, we have
\begin{equation}
\begin{aligned}
  \sideset{_{L}}{_{L}}{\mathop{\big\langle{0}|\mathclose X_{L}^{\mu}(\bar{z}_{i})  X_{L}^{\nu}(\bar{z}_{j})\mathclose  |{0}\big\rangle}}= {{{}_{}}_{}}_{L}\big\langle{0}| \Big(i\sqrt{\alpha' \over 2}\sum_{n>0}\frac{1}{-n}\overline{\alpha}^{\mu}_{-n}\bar{z}_{i}^{n}\Big) \Big(i\sqrt{\alpha' \over 2}\sum_{n>0}\frac{1}{m}\overline{\alpha}^{\nu}_{m}\bar{z}_{j}^{-m}\Big)|{0}\big \rangle_{L}\,,
  	\end{aligned}\label{}
\end{equation}
where we tacitly assumed the forward time ordering for the left-moving sector.
In this case, using the canonical commutation relation (\ref{cr2}) and the new vacuum (\ref{wrongvac}), we see that the left-moving sector yields the contribution
\begin{equation}
\frac{\alpha'}{2}\sum_{m,n>0}\frac{1}{mn}\sideset{_{L}}{_{L}}{\mathop{\left\langle 0|\overline{\alpha}^{\mu}_{-n}\overline{\alpha}^{\nu}_{m}|0\right\rangle}}\frac{\bar{z}^{n}_{i}}{\bar{z}^{m}_{j}} = -\eta^{\mu\nu}\frac{\alpha'}{2} \sum_{n>0} \frac{1}{n}\Big(\frac{\bar{z}_i}{\bar{z}_j}\Big)^{n}\sideset{_{L}}{_{L}}{\mathop{\left\langle 0|0\right\rangle}}\,.
\end{equation}
One finds that the left-moving sector converges only if the ordering is taken $|z_i| < |z_j|$. However, this backward ordering just resides outside the convergence range $|z_{i}|>|z_{j}|$ of the right-moving sector. We conclude that, if forward time ordering is taken for the left-moving sector as well as for the right-moving sector, the two-point correlations in the new vacuum is ill-defined because of lack of convergence. 

\subsection{Backward time-ordering for the new vacuum}
The problem posed above suggests a way out in itself: it must be that the lack of convergence came about because one took inconsistent time ordering that does not hold for the new vacuum (\ref{wrongvacuum})  \footnote{See \cite{Polchinski:1998rq} for a discussion about the relation between the time ordering and the normal ordering.}. Therefore, we now consider backward time ordering for the left-moving oscillators\ , viz.
\be
T_L\big[A(\overline{z_1})B(\overline{z_2})\big]=B(\overline{z_2})A(\overline{z_1})\,,\quad\quad \text{if} \ \ \ \vert z_1\vert>\vert z_2\vert \label{torder}\ .
\ee

Using the prescription (\ref{torder}), we now find that the contribution of left-moving oscillator is given by 
\begin{equation}
\begin{aligned}
  \sideset{_{L}}{_{L}}{\mathop{\big\langle{0}|T_{L}\big[X_{L}^{\mu}(\overline{z}_{i})  X_{L}^{\nu}(\overline{z}_{j})\big] |{0}\big\rangle}}&= {{{}_{}}_{}}_{L}\big\langle{0}| \Big(i\sqrt{\frac{\alpha'}{2}}\sum_{n>0}\frac{1}{-n}\overline{\alpha}^{\mu}_{-n}\overline{z}_{j}^{n}\Big) \Big(i\sqrt{\frac{\alpha'}{2}}\sum_{n>0}\frac{1}{m}\overline{\alpha}^{\nu}_{m}\overline{z}_{i}^{-m}\Big)|{0}\big \rangle_{L}\,,
  \\
  & = -\eta^{\mu\nu}\frac{\alpha'}{2} \sum_{n>0} \frac{1}{n}\Big(\frac{\overline{z}_j}{\overline{z}_i}\Big)^{n} \sideset{_{L}}{_{L}}{\mathop{\left\langle 0|0\right\rangle}}\,,
  \\
  & = + \eta^{\mu\nu}\frac{\alpha'}{2} \log\big(1-\frac{\overline{z}_j}{\overline{z}_i}\big)\sideset{_{L}}{_{L}}{\mathop{\left\langle 0|0\right\rangle}}\,, 
\end{aligned}\label{left_correlator}
\end{equation}
and that the resummation is well-defined provided the ordering is arranged to $|z_{i}|>|z_{j}|$. Note that the contribution of left-moving oscillators (\ref{left_correlator}) has opposite sign to the contribution of right-moving oscillators (\ref{right_correlator}), if the norms of the left-moving ground state and right-moving ground state had the same sign. 

It now remains to determine the zero-mode vacuum state. We do this by 
requiring that the two-point correlation function (\ref{2pointcorrelator}) is translation invariant, viz. it is a function of $z_{i}-z_{j}$ and $\overline{z}_{i}-\overline{z}_{j}$. As the time ordering we prescribed for the left-moving and right-moving oscillators are opposite each other, we shall also separate the center of mass position and momentum into left-moving and right-moving sectors, 
\begin{equation}
\begin{aligned}
  X^{\mu}_{0} &= X^{\mu}_{0_{R}} + X^{\mu}_{0_{L}}\,,
  \\
  P^{\mu} &= \tfrac{1}{2} \big(P^{\mu}_{R}+P^{\mu}_{L}\big)\,,
\end{aligned}\label{}
\end{equation}
where  $P^{\mu}_{L}$ and $P^{\mu}_{R}$ ($X^{\mu}_{0_{R}}$ and $X^{\mu}_{0_{L}}$) are treated as independent operators acting on mutually independent left-moving and right-moving Hilbert spaces, respectively. We then impose the commutation relation for the zero-mode sector
\begin{equation}
\begin{aligned}
  &\left[X^{\mu}_{0_{L}}, P^{\nu}_{L}\right] = i \eta^{\mu\nu}\,, &\qquad &\left[X^{\mu}_{0_{R}}, P^{\nu}_{R}\right] = i \eta^{\mu\nu}\,,
  \\
  &\left[X^{\mu}_{0_{L}},P^{\nu}_{R}\right] = 0\,,& \qquad &\left[X^{\mu}_{0_{R}}, P^{\nu}_{L}\right] =0\,.
\end{aligned}\label{zero_commutation}
\end{equation}
Note that we are tacitly assuming that our quantization scheme handles the left-moving and right-moving zero-modes separately even in the flat Minkowski spacetime.

Under the above assumption, the zero-mode vacuum would be divided into left-moving and right-moving sectors as well,
\begin{equation}
  |0\rangle_{0} \to |0\rangle_{0_{L}} \otimes |0\rangle_{0_{R}}\,.
\label{}\end{equation}
First, let us define  the zero-mode vacuum as
\begin{equation}
  P^\mu_{R}|0\rangle_{0_{R}} = 0\,, \qquad \sideset{_{0_{L}\!\!}}{}{\mathop{\langle}}0| P^\mu_{L} = 0\,.
\label{important}\end{equation}
Then using the zero-mode vacuum prescription (\ref{important}) and the backward time-ordering (\ref{torder}) for the left-moving sector, we find that the zero-mode contribution in the last line of (\ref{correlator1}) is replaced by
\begin{multline}
  {{{}_{}}_{}}_{0}\big\langle{0}| \Big(X^{\mu}_{0}-i\frac{\alpha'}{2}P^{\mu}\log\big({z}_{i} \overline{z}_{i}\big)\Big)  \Big(X^{\mu}_{0}-i\frac{\alpha'}{2}P^{\mu} \log\big({z}_{j} \overline{z}_{j}\big)\Big)|{0}\big \rangle_{0}
  \\
  \longrightarrow -i\frac{\alpha'}{2} \Big({{{{{}_{}}_{}}_{}}_{}}_{0_{R}\!\!}\big\langle{0}| P^{\mu}_{R} X_{0_{R}}^{\mu} \log z_{i}|0 \big \rangle_{0_{R}} + {{{{{}_{}}_{}}_{}}_{}}_{0_{L}\!\!}\big\langle{0}| X_{0_{L}}^{\mu}P^{\mu}_{L} \log \overline{z}_{i}|0 \big \rangle_{0_{L}}\Big)\,. 
\end{multline}
Here, we omit correlator between $X_{0_{R}}$ and $X_{0_{L}}$,  which amounts to an infrared regularization on the worldsheet.   
From the commutation relation of zero-modes in (\ref{zero_commutation}), we get
\begin{equation}
  -\frac{\alpha'}{2} \eta^{\mu\nu} \big(\log z_{i} - \log\overline{z}_{i} \big) = -\frac{\alpha'}{2} \eta^{\mu\nu} \log\frac{z_{i}}{\overline{z}_i}\,.
\label{zero_correlator}\end{equation}
Note that the zero-mode vacuum states, for which ${}_{0 _{L,R}} \langle 0 \vert P^\mu_{ L,R} \vert 0 \rangle_{0_{L,R}} = 0 $, are 
not normalizable, and so ${}_{0_{L,R}} \langle 0 \vert$ are not connected to $\vert 0 \rangle_{0_{L,R}}$ by Hermitian conjugation, $({}_{0_{L,R}} \langle 0 \vert)^\dagger \ne \vert 0 \rangle_{0_{L,R}}$. Accordingly, $P^\mu_{L, R} \vert 0 \rangle_{0_{L,R}}=0$ and ${}_{0_{L,R}} \langle 0 \vert P^\mu_{L, R}=0$ are not equivalent. So, while the operators of left-moving zero modes and of right-moving zero-modes obey identical commutation relations, their actions on the vacuum states come always with twofold options. In defining the new vacuum, we chose the asymmetric option (\ref{important}). The result (\ref{zero_correlator}) is then direct consequence of this choice.

We also note that the new vacuum (\ref{wrongvacuum}) can be expressed in a more compact and symmetric fashion,
\bea
\alpha_n^{\mu} \vert 0 \rangle = 0, \qquad \langle 0 \vert \overline{\alpha}_{n}^{\mu} \ = 0\,, \qquad n \geq0 \label{wrongvac1}\ .
\eea
Putting the contributions (\ref{right_correlator}), (\ref{left_correlator}) and (\ref{zero_correlator}) together, we finally have
\begin{equation}
\begin{aligned}
  \big\langle 0\big|T\big[X^{\mu}(z_{i},\overline{z}_{i})& X^{\nu}(z_{j},\overline{z}_{j})\big]\big|0\big\rangle 
  \\
  & = -\frac{\alpha'}{2}\eta^{\mu\nu}\log(\frac{z_i}{\overline{z}_i}) -\frac{\alpha'}{2}\eta^{\mu\nu}\Big(\log\big(1-\frac{z_j}{z_i}\big) - \log\big(1-\frac{\overline{z}_j}{\overline{z}_i}\big)  \Big)
  \\
  & = -\frac{\alpha'}{2}\eta^{\mu\nu}\log\Big(\frac{z_i-z_j}{\overline{z}_i-\overline{z}_j}\Big)\ .
\end{aligned}\label{two_point_ft}
\end{equation}
As proclaimed, the correlator is manifestly translation invariant. 
An identical result was obtained in \cite{Huang:2016bdd} but from a completely different consideration. 

\vskip0.5cm
We now summarize our quantization scheme over the new vacuum. 
\vskip0.5cm
\begin{tcolorbox}
\begin{itemize}
\vskip0.5cm
  \item equal time canonical commutation relations:
\be
[X^{\mu}(\tau,\sigma), \Pi^{\nu}(\tau,\sigma^{\prime})]=i\ \eta^{\mu\nu}\delta(\sigma-\sigma^{\prime})\,,
\ee
\begin{equation}
\begin{aligned}
  &\big[\overline \alpha^\mu_m, \overline \alpha^\nu_{n}] = m \, \delta_{m+n,0}\eta^{\mu \nu}\,, & &\big[\alpha^\mu_m, \alpha^\nu_{n}\big] =m \, \delta_{m+n,0} \eta^{\mu \nu} \,,
  \\
  &\big[X^{\mu}_{0_{L}}, P^{\nu}_{L}\big] = i \eta^{\mu\nu}\,, & &\big[X^{\mu}_{0_{R}}, P^{\nu}_{R}\big] = i\eta^{\mu\nu}\,.
\end{aligned}\label{}
\end{equation}
\vskip0.5cm
  \item new vacuum:
\begin{equation}
  	\left.\begin{aligned}\alpha_{n} |0\rangle_R &= 0\,,\qquad & {}_L\langle0|\overline{\alpha}_{n} &= 0\,,
  	\end{aligned}\right.\qquad \text{for}~ n\geq0\,.
\label{}\end{equation}
\vskip0.5cm
  \item time ordering:
\begin{equation}
\left.\begin{aligned} T_{R}\big[B(z_{j}) A(z_{i}) \big] &= A(z_{i}) B(z_{j})\,,
\\ T_{L}\big[\overline{A}(\overline{z}_{i}) \overline{B}(\overline{z}_{j}) \big] &= \overline{B}(\overline{z}_{j}) \overline{A}(\overline{z}_{i})\,, \end{aligned}\right.\qquad \text{for}~ |z_{i}|>|z_{j}|\,.
\end{equation}
\end{itemize}
\vskip0.5cm
\end{tcolorbox}


\vskip0.5cm
\subsection{The spectrum} \label{spectrum}

We now study the representation theory of Virasoro algebra over the new vacuum and extract the spectrum. First, we define the level operators that are compatible with (\ref{vc1}). The right-moving sector is defined as usual, but the left-moving sector is defined as \footnote{Note that we could rename the left level operator without the minus sign in front,  i.e. 
\be
{\overline{\bf N}}  =  \sum_{n=1}^{\infty}\overline{\alpha}_{n}\cdot\overline{\alpha}_{-n}\ \nonumber\ ,
\ee
however, in this case the eigenvalues of this operator would be negative definite,
\be
{\overline{\bf N}}| \overline N_n\rangle=-n\overline N_n | \overline N_n\rangle  \nonumber 
\ee
where $| \overline N_n\rangle$ is a base on the Hilbert space of left-moving sector.}
\begin{equation}
\begin{aligned}
  \bf N &= + \sum_{n=1}^{\infty}:\alpha_{-n}\cdot\alpha_{n}:\,,
\\
{\overline{\bf N}} &= -\sum_{n=1}^{\infty} :\overline{\alpha}_{n}\cdot\overline{\alpha}_{-n}:\,.
\end{aligned}\label{number_def}
\end{equation}
Using the level operators, the Virasoro operators are expressed as
\bea\label{viraope}
L_0 & = &\frac{1}{2}\alpha_0^2+ {\bf N}-a\,, \\ \nonumber
\overline{L}_0 & = &\frac{1}{2}\overline{\alpha}_0^2- {\overline{\bf N}}-\overline a\ ,
\eea
where $\alpha^\mu_{0} = \sqrt{\alpha' \over 2} P^\mu_{R}$ and $\overline{\alpha}^\mu_{0} = \sqrt{\alpha' \over 2} P^\mu_{L}$. The constants $a, \bar a$ depend on the choice of vacuum. 

The Virasoro conditions for a physical state $ | \mbox{phys} \rangle$ of closed string are then
\be \label{matrixelementsPT}
\langle \text{phys} | T_{ab} | \text{phys} \rangle  = 0\ .
\ee
We suppose that $L_0$ acts to the right, but, unlike conventional string, $\overline{L}_0$ acts to the left because of the presence of $P^\mu_{L}$ in it. Then, the physical conditions should be imposed on the full matrix elements (\ref{matrixelementsPT}). For instance,  
\be
\langle \text{phys} | L_0 | \text{phys}\rangle \pm\langle \text{phys} | \overline{L}_0 | \text{phys}\rangle=0
\label{phy0}\ .
\ee 
These imply the level-matching constraint
\be
 {\bf N}+\overline{\bf N}=a-\overline a \,,
\ee
and the mass-shell condition, $p_\mu p^\mu = -M^2$, with
\be
M^2=\frac{4}{\alpha^{\prime}}({\bf N}-a)=\frac{4}{\alpha^{\prime}}(-\overline{\bf N}-\overline a)\ .
\ee

The normal ordering constants $a$ and $\overline a$, which were left undetermined so far, can be fixed in several ways.  Here, we adopt a short-cut argument similar to what we described above for the conventional string theory, and relegate a rigorous proof based on the BRST formulation to the next sub-section.
The idea is that we demand that the state $\alpha^\mu_{-1}\overline{\alpha}^\nu_{+1}|0\rangle$ would give rise to `massless' string gravity excitations. This way, we find that $a=1$ and $\overline a=-1$. Then, the level-matching and the mass-shell conditions can be written as
\be
{\bf N}+{\overline{\bf N}}=2\,,
\ee
and
\be
M^2=\frac{4}{\alpha^{\prime}}({\bf N}-1)=\frac{4}{\alpha^{\prime}}(-{\overline{\bf N}}+1)\,. \label{massformulaa}
\ee
We immediately see that the spectrum satisfying zero-mode Virasoro conditions consists of a finite number of states, given in Table \ref{table1}:
\begin{table}[ht]
\centering
\begin{tabular}{|c|c|c|c|}
\hline\hline
${\bf N}$ & ${\overline{\bf N}}$ & $M^2$ & state \\
\hline
1 & 1 & 0 & $\epsilon_{\mu\nu}\alpha_{-1}^{\mu}\overline{\alpha}_{+1}^{\nu}|0,k\rangle$ \\
\hline
2 & 0 & $+\frac{4}{\alpha^{\prime}}$  &$ a_{\mu\nu} \alpha_{-1}^{\mu}\alpha_{-1}^{\nu}|0,k\rangle\oplus a_{\mu}\alpha_{-2}^{\mu}|0,k\rangle$ \\
\hline
0 & 2 & $-\frac{4}{\alpha^{\prime}}$ & $\overline{a}_{\mu\nu}\overline{\alpha}_{+1}^{\mu}\overline{\alpha}_{+1}^{\nu}|0,k\rangle\oplus \overline{a}_{\mu}\overline{\alpha}_{+2}^{\mu}|0,k\rangle$ \\
\hline\hline
\end{tabular}
\caption{Spectrum satisfying the zero-mode Virasoro conditions. $\epsilon_{\mu \nu}$, $a_{\mu \nu}$, $a_\mu$, $\bar a_{\mu \nu}$ and $\bar a_\mu$ are polarization tensors of the respective states.}{\label{table1}}
\end{table}

Additionally, we need to demand the harmonic Virasoro conditions 
\be
\langle \text{phys}| L_m | \text{phys}\rangle=0\,,\qquad\qquad \langle \text{phys}| \overline{L}_m | \text{phys}\rangle=0 \label{phys}\ .
\ee
In usual string theory, it is sufficient to demand  $L_m | \text{phys}\rangle=0$ and $\overline{L}_m | \text{phys}\rangle=0$ for all $m>0$. For us,  due to the choice of new vacuum,  we have to demand conditions as 
\bea \label{phys1}
L_m | \text{phys}\rangle & = & 0 \qquad\qquad m>0\,, \\ \nonumber
\langle \text{phys} |\overline{L}_m  & = & 0 \qquad\qquad m>0 \ , 
\eea
which are in fact still compatible with the conditions (\ref{phys}).
Now, using Eq. (\ref{phys1}), we get the string physical spectrum given in Table \ref{table3}.
\begin{table}[ht]
\centering
\begin{tabular}{|c|c|c|c|c|c|}
\hline\hline
${\bf N}$ & ${\overline{\bf N}}$ & $M^2$ & state & gauge condition & norm \\
\hline
1 & 1 & 0 & $\epsilon_{\mu\nu}\alpha_{-1}^{\mu}\overline{\alpha}_{+1}^{\nu}|0,k\rangle$ & $k^{\mu}\epsilon_{\mu\nu}=k^{\nu}\epsilon_{\mu\nu}=0$ & +1 \\
\hline
2 & 0 & $+\frac{4}{\alpha^{\prime}}$  & $ a_{\mu\nu} \alpha_{-1}^{\mu}\alpha_{-1}^{\nu}|0,k\rangle$ & $k^{\mu}a_{\mu\nu}= a_{\mu}{}^{\mu}=0$ & $-1$ \\
\hline
0 & 2 & $-\frac{4}{\alpha^{\prime}}$ & $\overline{a}_{\mu\nu}\overline{\alpha}_{+1}^{\mu}\overline{\alpha}_{+1}^{\nu}|0,k\rangle$ & $k^{\mu}\overline{a}_{\mu\nu}= \overline{a}_{\mu}{}^{\mu}=0$ & $-1$\\
\hline\hline
\end{tabular}
\caption{Spectrum satisfying the harmonic Virasoro conditions. Their norms are tabulated in the last column.}{\label{table3}}
\end{table}
\hfill\break
The first state in Table \ref{table3} contains the graviton, the Kalb-Ramond field and the dilaton, containing $(D-2)^2$ degrees of freedom as in the conventional string theory. The other two states are both massive Pauli-Fierz spin-two particles, obeying the on-shell conditions 
\be
k^{\mu}a_{\mu\nu}= a_{\mu}{}^{\mu}=0\label{separateequ}\ , 
\qquad 
k^{\mu} \overline a_{\mu\nu}= \overline a_{\mu}{}^{\mu}=0 \ , 
\ee
and comprise of $\frac{1}{2}(D-2)(D+1)$ degrees of freedom.

In the last column of Table \ref{table3}, we also tabulated the norm of each state. In fact, due to our normal-ordering convention, the norm has a two-fold ambiguity. Computing the norm of these three states using the commutation relation and (\ref{wrongvac}), we get
\begin{equation}
\begin{aligned}
  &\left\langle 0,k|\alpha_{+1} \overline{\alpha}_{-1} \alpha_{-1} \overline{\alpha}_{+1} |0,k\right\rangle \sim - \left\langle 0|0\right\rangle\,,
  \\
  &\left\langle 0,k|\alpha_{+1} \alpha_{+1} \alpha_{-1} \alpha_{-1} |0,k\right\rangle \sim + \left\langle 0|0\right\rangle\,,
  \\
  &\left\langle 0,k|\overline{\alpha}_{-1} \overline{\alpha}_{-1} \overline{\alpha}_{+1} \overline{\alpha_{+1}} |0,k\right\rangle \sim  +\left\langle 0|0\right\rangle\,.
\end{aligned}\label{}
\end{equation}
If the vacuum (which is not part of physical state) were a positive norm state, $\langle 0|0\rangle>0\,,$ then the gravity multiplet would have been negative norm states. The resolution to this trouble is provided by assigning a negative norm to the zero-mode vacuum  state,
\be
\langle 0|0\rangle<0 \ .
\ee
This then render massive modes to be negative-norm states. See Table \ref{table3}.

Let us summarize where we are. By quantizing closed string on the new vacuum we propose, we obtained only a finite number of states. They are
\begin{tcolorbox}
\leftline{\bf the string spectrum}
\begin{itemize}
\item string gravity states: they comprise of spacetime metric $g_{\mu \nu}$, Kalb-Ramond field $B_{\mu \nu}$, and dilaton field $\phi$. They are massless and all have positive-norm. 
\item a pair of excited states : these comprise of spin-two  
states $a_{\mu \nu}$ and $\overline a_{\mu \nu}$. They have mass-squared $\pm 4/ \alpha'$ and both negative-norm. 
\end{itemize}
\end{tcolorbox}
\noindent
The field content in Table \ref{table3} is exactly the same as the spectrum of HSY prescription \cite{Huang:2016bdd}. The difference is that, while the HSY prescription obtains this spectrum from a certain modified rule of the KLT relation, we obtain the same spectrum from the {\sl ab initio} standard canonical quantization of string theory over the new vacuum. 

While we have built the string spectrum by imposing the Virasoro constraints for the zero modes (\ref{phy0}) as well as the harmonic modes (\ref{phys1}), it would also be interesting to compare the field contents at intermediate steps. Indeed, the spectrum of Table \ref{table1} is obtained by imposing the zero modes of Virasoro constraints (\ref{phy0}) only. It might be regarded as the field contents from a version of massive gravity. 
In particular, up to Table \ref{table1}, the massive ghost fields do not have to satisfy the Virasoro constraints for harmonic modes (\ref{phys1}).  Interestingly, they resemble the massive fields that are present in the $\alpha'$-corrected double field theory of HSZ \cite{Hohm:2016lim}. Further imposing the harmonic modes (\ref{phys1}), we obtained the spectrum of Table \ref{table3}. As such, the field contents in Table \ref{table1} are larger than those in Table \ref{table3}: the spectrum in Table \ref{table1} contains longitudinal vector and scalar components to the massive spin-two states.  
  
This observation leads us to two viable interpretations of the $\alpha'$-corrected double field theory of HSZ. On one hand, the spectrum of HSZ may be regarded the same as the spectrum in Table {\ref{table1}}. Indeed, HSZ imposed the Virasoro constraints only for the zero modes. On the other hand, it could be the case that the field theory action proposed in \cite{Hohm:2016lim} is not complete. The completion would generate extra terms in the Lagrangian, which would then render further on-shell conditions (\ref{separateequ}). Indeed, this is what happens in the context of massive gravity.


\subsection{The $b, c$ ghosts and the normal ordering constants}
We can also proceed with the covariant quantization. In this case, it is necessary to introduce the $b$, $c$ ghost fields as part of the worldsheet degrees of freedom. With all the subtleties associated with the new vacuum understood for the chiral bosons $X_{L, R}^\mu$, the steps leading to the covariant quantization can be repeated straightforwardly. As such, here we will focus on determining the intercept constants $a$ and $\bar a$ for the new vacuum. The discussion is essentially parallel to that in the conventional string theory. The main difference is that, unlike in the conventional string theory, as here we are choosing a new vacuum configuration for the left-moving sector, we also need to choose a new vacuum for the left-moving sector of $b$, $c$ ghosts. 

Including the contribution of $b, c$ ghosts, the total Virasoro generators are
\bea \label{virasoro_bc}
L^{\rm tot}_m & = & \frac{1}{2}\sum_n\,\mathclose:\alpha_{m-n}\alpha_{n}\mathclose:+\sum_n(2m-n)\mathclose:b_{n}c_{m-n}\mathclose:\\ \nonumber
\bar{L}^{\rm tot}_m & = & \frac{1}{2}\sum_n\,\mathclose:\overline{\alpha}_{m-n}\overline{\alpha}_{n}\mathclose:+\sum_n(2m-n)\mathclose:\overline{b}_{n}\overline{c}_{m-n}\mathclose: \qquad (m \ne 0)
\eea
for nonzero modes, and 
\bea\label{virasoro_bc_0}
L^{\rm tot}_0       & = & L_0-\sum_n n\mathclose:b_{n}c_{-n}\mathclose: - a \\ \nonumber
\bar{L}^{\rm tot}_0 & = & \bar{L}_0-\sum_n n \mathclose:\overline{b}_{n}\overline{c}_{-n}\mathclose: - \overline{a} \ 
\eea
for zero modes. They satisfy the Virasoro algebra without the central extension
\bea\label{vira_bc}
\Big[L^{\rm tot}_m, L^{\rm tot}_n\Big] & = & (m-n)L^{\rm tot}_{m+n}\,,
\\ \nonumber
\Big[\overline{L}^{\rm tot}_m, \overline{L}^{\rm tot}_n\Big] & = & (m-n)\overline{L}^{\rm tot}_{m+n}\ .
\eea

The new vacuum for the ghost fields can be defined completely parallel. For the right-moving oscillators,  we prescribe conventionally
\be
c_m | 0 \rangle=0 \qquad m>0\,,
\ee
\be
b_m | 0 \rangle=0 \qquad m>0\,,
\ee
while for the left-moving oscillators, we prescribe as 
\be
\langle 0 |\overline{c}_m =0 \qquad m>0\,,
\ee
\be
\langle 0 | \overline{b}_m=0  \qquad m>0\ .
\ee
The physical conditions for harmonic modes are given by 
\be
\langle \text{phys} | L^{\rm tot}_m | \text{phys} \rangle=0 \qquad m>0\,,
\ee
\be
\langle \text{phys} | \bar{L}^{\rm tot}_m  | \text{phys} \rangle = 0 \qquad m<0\,. 
\ee
For the zero modes, for both the right-moving sector and the left-moving sector, we prescribe conventionally,
\bea
b_0| \uparrow \rangle =   | \downarrow \rangle\,, &\qquad & b_0| \downarrow \rangle= 0\,,
\\ \nonumber
c_0| \uparrow \rangle  =   0 \,, &\qquad &   c_0| \downarrow \rangle= | \uparrow \rangle\ , 
\eea
\bea
\overline b_0\overline{| \uparrow \rangle} =   \overline{| \downarrow \rangle}\,, & \qquad  & \overline b_0\overline{| \downarrow \rangle}= 0\,, \\ \nonumber
\overline c_0\overline{| \uparrow \rangle}  =   0 \,, & \qquad&   \overline c_0\overline{| \downarrow \rangle}=\overline{ | \uparrow \rangle}\ .
\eea

With all these, it is straightforward to compute the normal ordering constants.  Using the Virasoro algebra (\ref{vira_bc}) for the particular values  $m=1$ and $n=-1$,  we get 
\be
\Big[L^{\rm tot}_1, L^{\rm tot}_{-1}\Big]=2L^{\rm tot}_{0}\,,
\ee
\be
\Big[\bar{L}^{\rm tot}_1, \bar{L}^{\rm tot}_{-1}\Big]=2\bar{L}^{\rm tot}_{0}\ .
\ee
Then, using the physical conditions, we can compute $L^{\rm tot}_0| 0 \rangle$ and  $\langle 0 | \bar{L}^{\rm tot}_0$ through the above commutation relations,  viz.
\bea
L^{\rm tot}_0| 0 \rangle & = & +\frac{1}{2} L^{\rm tot}_1L^{\rm t}_{-1}| 0 \rangle\,,
\\ \nonumber
\langle 0 |\overline{L}^{\rm tot}_0 & = & -\frac{1}{2}\langle 0 | \overline{L}^{\rm tot}_{-1}\overline{L}^{\rm tot}_{1} .
\eea
By comparing the final result,
\bea
L^{\rm tot}_0| 0 \rangle=-| 0 \rangle\,,\\ \nonumber
\langle 0 |\overline{L}^{\rm tot}_0=+\langle 0 |\,,
\eea
with the definition of the Virasoro operators (\ref{virasoro_bc_0}), we conclude that 
\be
a=-1\,, \qquad\qquad \overline{a}=1\ .
\ee
On the other hand, in the unitary, light-cone gauge, these constants can be related to  the spacetime dimension, 
\bea\nonumber
a & = & \frac{2-D}{24}\\
\overline{a} & = & \frac{D-2}{24} \ , 
\eea
The Lorentz invariance asserts that the unitary gauge and the covariant gauge should yield identical result. This leads to the conclusion that, as in the conventional bosonic string theory, the critical dimension is $D=26$.


\section{Tree-level scattering amplitude}\label{amplitudes}

In this section, we study interactions of the closed bosonic string over the new vacuum.  To be specific, we shall compute the tree-level scattering amplitude of four dilatons in the operator formalism. This is the simplest amplitude, but still displays several important features pertaining to the theory. We emphasize that our results are computed in the standard string theory framework; the only change now is the vacuum choice. In the course of this computation, we will also clarify limitations of the HSY prescription that flips the sign of spacetime metric when specifically applied to the dilaton four-point scattering amplitude.

\subsection{Correlation functions of master vertex operators}
We first introduce master vertex operator $\cV(k,z,\overline{z};\xi,\overline{\xi})$, which provides an efficient method for computing scattering amplitudes of arbitrary higher excitation modes,
\begin{equation}
 \cV(k,z,\overline{z},\xi,\overline{\xi}) = :\exp[ik\cdot X(z,\overline{z}) + i \xi\cdot \partial X_{R}(z) + i \overline{\xi}\cdot \overline{\partial}X_{L}(\overline{z})] :\,.
\label{mVertex}\end{equation}
One finds it convenient to decompose $\cV(k,z,\overline{z};\xi,\overline{\xi})$ into the right-moving and left-moving operators, 
\bea
\cV_R(k,z) & = & \exp(i k\cdot X_R+i\xi\cdot\partial X_R )\,,\\ \nonumber
\cV_L(k,\overline{z}) & = & \exp(i k\cdot X_L+i\overline{\xi}\cdot\overline{\partial} X_L )\ .
\eea
Then, the $M$-points function of $\cV(k,z,\overline{z};\xi,\overline{\xi})$ is also factorized into two parts
\begin{equation}
\begin{aligned}
  \cA_{M}(1,2,\cdots,M) & =  \left\langle T\big[\cV(1)\cV(2)\ldots \cV(M)\big]\right\rangle\\
{} & = \left\langle T_R\big[ \cV_R(k_1,z_1)\cdots \cV_R(k_M,z_M)\big]\right\rangle_R \left\langle  T_L\big[\cV_L(k_1,\overline{z}_1)\cdots \cV_L(k_M,\overline{z}_M)\big]\right\rangle_L \,,
\end{aligned}\label{}
\end{equation}
where $T_{R}$ stands for forward time-ordering, while $T_{L}$ stands for backward time-ordering, as defined in (\ref{torder}).
For the right-moving sector, we get the same result as in the conventional string theory. Here, we present the result without providing details of the computation \footnote{ See \cite{Green:1987sp} for the detailed computation using the operator formalism .}
\begin{equation}
\begin{aligned}
   \big\langle T_R\big[ &\cV_R(k_1,z_1)\cdots \cV_R(k_M,z_M)\big]\big\rangle 
   \\
   &=\prod_{i<j}(z_i-z_j)^{ \frac{\alpha'}{2}k_i\cdot k_j}\exp\Big[\frac{\alpha'}{2}\sum_{i<j}\Big(\frac{\xi_i\cdot\xi_j}{(z_i-z_j)^2}+\frac{\xi_i\cdot k_j}{(z_i-z_j)}-\frac{k_i\cdot\xi_j}{(z_i-z_j)}\Big)\Big] \ .	
\end{aligned}\label{rightcontri}
\end{equation}

The computation for the left-moving sector is trickier, so we shall present some details of the computation. We first rewrite the left-moving vertex operator in the following form,
\be
\cV_L(k,\overline{z})=\exp\Big(i k\cdot X_{0L}+\frac{\alpha'}{2}f(\overline{z})\cdot P_L+\sqrt{\frac{\alpha'}{2}}\sum_{n\neq0}g_n(\overline{z})\cdot\overline{\alpha}_n\Big)\ ,
\ee
where
\bea
f^{\mu}(\overline{z}) & = & k^{\mu}\ \log(\overline{z})+\frac{\overline{\xi}^{\mu}}{\overline{z}}\\ \nonumber
g_n^{\mu}(\overline{z}) & = & -\frac{1}{n}k^{\mu}\ \overline{z}^{-n}+\overline{\xi}^{\mu}\ \overline{z}^{-n-1}\ .
\eea
Next, we split the vertex operator into the zero mode part and the oscillator part
\be
V_L(k,\overline{z})=\exp\Big(i k\cdot X_{0L}+\frac{\alpha'}{2}f(\overline{z})\cdot P_L\Big)\ \exp\Big(\sqrt{\frac{\alpha'}{2}}\sum_{n\neq0}g_n(\overline{z})\cdot\overline{\alpha}_n\Big):=Z_0Z_L\ .
\ee
As the operators $Z_0$ and $Z_L$ act on different Fock spaces, the left-moving contribution to the amplitude can also be split  into two parts :
\be
A_{0L}A_L={}_{0L}\langle 0\vert Z_0(k_1,\overline{z}_1)\ldots Z_0(k_M,\overline{z}_M)\vert 0 \rangle_{0L}\ \cdot \ {}_L\langle 0\vert Z_L(k_1,\overline{z}_1)\ldots Z_L(k_M,\overline{z}_M)\vert 0 \rangle_L\ .
\ee
To compute the zero mode contribution, as we need to use (\ref{important}),  we move the $P_L$ operators to the left. Using the Baker-Campbell-Haussdorf formula
$\text{e}^{A+B}=\text{e}^B\  \text{e}^A \ \text{e}^{\frac{1}{2}[A,B]}
$,  
it can be written as
\be
Z_0(k,\overline{z})=\exp\big(\frac{\alpha'}{2}f(\overline{z})\cdot P_L\big)\ \exp\big(i k\cdot X_{0L}\big)\ \exp\big(-\frac{\alpha'}{4}f(\overline{z})\cdot k\big)\ .\label{z0}
\ee
Now, taking into account that $P_L$ acts on the left and that $\langle -k \vert=\langle 0 \vert  \exp\big(i k\cdot X_{0L}\big)$, after some algebra, one gets
\be
A_{0L}(k)=\exp\Big( -\frac{\alpha'}{2}\sum_{i<j}(k_i\cdot k_j\ \log(\overline{z}_i)-\frac{\alpha'}{2}\frac{\overline{\xi}_i\cdot k_j}{\bar{z_i}}) \Big)\ .
\ee

The calculation for the left-moving oscillators proceeds analogous to the calculation for the right-moving ones except for the backward time-ordering (\ref{torder}) on the left-moving oscillators. The coherent states method and the details of this computation, in the conventional case, can be found in \cite{Green:1987sp}. Taking into account of (\ref{torder}), it can be easily translated to this set up. After some calculation, one gets
\begin{equation}
\begin{aligned}
\big\langle  T_L\big[&\cV_L(k_1,\overline{z}_1)\cdots \cV_L(k_M,\overline{z}_M)\big]\big\rangle
   \\
   &=\prod_{i<j}(\overline{z}_i-\overline{z}_j)^{- \frac{\alpha'}{2}k_i\cdot k_j}\exp\Big[-\frac{\alpha'}{2}\sum_{i<j}\Big(\frac{\overline{\xi}_i\cdot\overline{\xi}_j}{(\overline{z}_i-\overline{z}_j)^2}+\frac{\overline{\xi}_i\cdot k_j}{(\overline{z}_i-\overline{z}_j)}-\frac{k_i\cdot\overline{\xi}_j}{(\overline{z}_i-\overline{z}_j)}\Big)\Big]\ .	
\end{aligned}\label{leftcontri}
\end{equation}
Note the minus sign in the exponent of (\ref{leftcontri}), in contrast to (\ref{rightcontri}).

The complete expression for the $M$-point scattering amplitude can be written as
\bea\label{fullA}
\cA_{M}(1,\cdots,M) & = & (2\pi)^D\ \delta^D\big(\sum_{i}k_i\big) \prod_{i<j}\Big(\frac{z_i-z_j}{\overline{z}_i-\overline{z}_j}\Big)^{ \frac{\alpha'}{2}k_i\cdot k_j}\\ \nonumber
{} & {} &\cdot \exp\Big[+\frac{\alpha'}{2}\sum_{i<j}\Big(\frac{\xi_i\cdot\xi_j}{(z_i-z_j)^2}+ \frac{\xi_i\cdot k_j}{(z_i-z_j)}- \frac{k_i\cdot\xi_j}{(z_i-z_j)}\Big)\Big] \\ \nonumber
{} & {} & \cdot \exp\Big[-\frac{\alpha'}{2}\sum_{i<j}\Big(\frac{\overline{\xi}_i\cdot\overline{\xi}_j}{(\overline{z}_i-\overline{z}_j)^2}+ \frac{\overline{\xi}_i\cdot k_j}{(\overline{z}_i-\overline{z}_j)}- \frac{k_i\cdot\overline{\xi}_j}{(\overline{z}_i-\overline{z}_j)}\Big)\Big] \ .
\eea
Here, the energy-momentum conservation comes from the integration of zero-mode part
\be
(2\pi)^D\ \delta\big(\sum_{i}k_i\big) =\int d^DX_{0L}d^DX_{0R}\, \delta\big({X_{0L}-X_{0R}}\big)e^{i  \sum_i k_{i}\cdot(X_{0L}+ X_{0R})}\label{xlxriden}\ ,
\ee
where we also imposed the identification between zero modes of $X_L$ and $X_R$.

\subsection{Four dilaton scattering amplitude}
We now apply the above master formula for computing the scattering amplitude of dilatons. The four dilaton scattering amplitude is given by the four-point correlation function of dilaton vertex operators
\begin{equation}
  M^{\scriptscriptstyle D}_{4}(1,2,3,4) = \int \dd \mu_4 \left\langle T\big[V_{D}(1) V_{D}(2) V_{D}(3) V_{D}(4) \big]\right\rangle\,. 
\label{ScatteringAmp}\end{equation}
By identical particle nature, the scattering amplitude ought to be permutation symmetric among the four dilaton quantum numbers. 
Here, the dilaton vertex operator $V_{D}$ is given by
\begin{equation}
  V_{D}(z,\overline{z}) = -\frac{2}{\alpha'}g_c\,\varepsilon^{\scriptscriptstyle (D)}_{\mu\nu}\  \partial X_R(z) ^{\mu}\exp( k\cdot X_R(z)) \ \overline{\partial} X_L(\overline{z})^{\nu}\exp( k\cdot X_L(\overline{z}))\,,
\label{}\end{equation}
where the momentum quantum number $k^\mu$ obeys the mass-shell condition $k^{2}=0$. The $\dd \mu_4$ is the integration measure which defines $SL(2,\mathbb{C})$ invariant amplitude, 
\begin{equation}
  \dd \mu_4 = \dd^2 z_{1}\dd^2 z_{2}\dd^2 z_{3}\dd^2 z_{4} |z_{1}-z_{2}|^2 |z_{1}-z_{4}|^2 |z_{2}-z_{4}|^2 \delta^2(z_1-z_1^{0})\delta^2(z_2-z_2^{0})\delta^2(z_4-z_4^{0})\,. 
\label{measure}\end{equation}
Note that (\ref{measure}) is equivalent to the conventional string case, as the $SL(2,\mathbb{C})$ conformal symmetry is still maintained by the choice of new vacuum. 
So, we can take the standard choice for the position of vertex operators such that $z_3 \to z$ is the moduli variable and the other three vertex operators are fixed at positions $z^{0}_{1}\to \infty$, $z^{0}_{2}= 1$, and $z^{0}_{4}=0$. 

The dilaton vertex operator can be obtained from (\ref{mVertex}) by taking derivatives with respect to the $\xi$ and $\overline{\xi}$
\begin{equation}
  V_{D}(z,\overline{z}) = -\frac{2}{\alpha'}g_c\,\varepsilon^{\scriptscriptstyle (D)}_{\mu\nu} \frac{\partial}{\partial \xi_{\mu}}\frac{\partial}{\partial \overline{\xi}_{\nu}} \cV(z,\overline{z};\xi,\overline{\xi})\Big|_{\xi=\overline{\xi}=0}\,,
\label{dilatonVertex}\end{equation}
where $\varepsilon^{\scriptscriptstyle (D)}_{\mu\nu}$ is the dilaton polarization tensor satisfying $k^{\mu}\varepsilon^{\scriptscriptstyle (D)}_{\mu\nu}=0$. It can be explicitly represented as
\begin{equation}
  \varepsilon^{\scriptscriptstyle (D)}_{\mu\nu} = \frac{1}{\sqrt{D-2}}(\eta_{\mu\nu}-k_{\mu}\overline{k}_{\nu}-\overline{k}_{\mu}k_{\nu})\,,
\label{}\end{equation}
where $\overline{k}_{\mu}$ is an auxiliary vector satisfying $\overline{k}^2=0$ and $\overline{k}\cdot k=1$. 

At this stage, one might like to understand how our {\sl ab initio} computation of the scattering amplitudes is compared to that within the HSY prescription of flipping the sign of target space metric. Here, we emphasize that the HSY prescription cannot be applied for computing the dilaton scattering amplitude. The polarization tensor of the general massless spin-2 field $\varepsilon_{\mu\nu}$ is decomposed into the  right-moving and left-moving sectors, which are represented by $\zeta_{\mu}$ and $\overline{\zeta}_{\mu}$ respectively,
\begin{equation}
  \varepsilon_{\mu\nu} = \zeta_{\mu} \overline{\zeta}_{\nu}\,.
\label{decomp_polarization}\end{equation}
The symmetric part $\varepsilon^{(S)}_{\mu\nu} = \zeta_{(\mu} \overline{\zeta}_{\nu)}$ contains the trace part corresponding to  the dilaton polarization,
\begin{equation}
  \frac{\zeta \cdot \overline{\zeta}}{D-2} \big(\eta_{\mu\nu} -k_{\mu}\overline{k}_{\nu}-\overline{k}_{\mu}k_{\nu}\big)\,.
\label{trace_part}\end{equation}
In order to apply the HSY prescription, the definite separation of left-moving and right-moving sectors is crucial. However, the trace part (\ref{trace_part}) has a contraction between the left-moving and right-moving sectors, $\zeta\cdot\overline{\zeta}$. Therefore an ambiguity arises  for choosing the metric between the metric for right-moving sector and the metric for left-moving sector, $\eta_{\mu\nu}$ and $\overline{\eta}_{\mu\nu} = - \eta_{\mu\nu}$, respectively. Thus, strictly speaking, the metric sign flipping prescription of HSY is not applicable. On the other hand, the canonical quantization over the new vacuum as we proceeds presently is {\sl ab initio} approach and hence fully applicable even for the dilaton scattering amplitude.

The four-point correlation function of $V_{D}$ can be straightforwardly computed from (\ref{fullA}) and (\ref{dilatonVertex}) by repeatedly taking derivatives with respect to $\xi$ and $\overline{\xi}$
\begin{equation}
\begin{aligned}
	\big\langle T\big[V_{D}(1) &V_{D}(2) V_{D}(3) V_{D}(4)\big] \big\rangle 
	\\
	&= \Big(\frac{2}{\alpha'}\Big)^4 g_c^4 \ \varepsilon^{\scriptscriptstyle (D)}_{1 \mu_1\nu_1} \cdots \varepsilon^{\scriptscriptstyle (D)}_{4 \mu_4\nu_4} \frac{\partial}{\partial \xi_{1\mu_1}} \frac{\partial}{\partial \overline{\xi}_{1\nu_1}} \cdots \frac{\partial}{\partial \xi_{4\mu_4}} \frac{\partial}{\partial \overline{\xi}_{4\mu_4}}\cA_{M}(k_i, z_i, \overline{z}_i,\xi_i,\overline{\xi}_i) |_{\xi_i = \overline{\xi}_i=0}\,.
\end{aligned}
\end{equation}
This expression can be divided into the left-moving and right-moving sectors \footnote{See Appendix \ref{4pointappendix} for explicit form of the $F_R$ and $F_L$.}
\begin{equation}
\begin{aligned}
\big\langle T\big[&V_{D}(1) V_{D}(2) V_{D}(3) V_{D}(4)\big] \big\rangle 
	\\
	&= \Big(\frac{2}{\alpha'}\Big)^4 g_ c^4 \ \varepsilon^{\scriptscriptstyle (D)}_{1 \mu_1\nu_1} \cdots \varepsilon^{\scriptscriptstyle (D)}_{4 \mu_4\nu_4} F_{R}{}^{\mu_1 \mu_2 \mu_3 \mu_4}(z_1,z_2,z_3,z_4) F_{L}{}^{\nu_1 \nu_2 \nu_3 \nu_4}(\overline{z}_1,\overline{z}_2,\overline{z}_3,\overline{z}_4)\,.
\end{aligned}
\label{}\end{equation}
The scattering amplitude (\ref{ScatteringAmp}) is reduced to 
\begin{equation}
  M_4^{\scriptscriptstyle D}(1,2,3,4) = \Big(\frac{2}{\alpha'}\Big)^4 g_c^4 \  \int \dd^2 z \varepsilon^{\scriptscriptstyle (D)}_{1 \mu_1\nu_1} \cdots \varepsilon^{\scriptscriptstyle (D)}_{4 \mu_4\nu_4} F_{R}{}^{\mu_1 \mu_2 \mu_3 \mu_4}(z) F_{L}{}^{\nu_1 \nu_2 \nu_3 \nu_4}(\overline{z})\ .
\label{ScatAmp}\end{equation}
The explicit forms of $F_{R}$ and $F_{L}$ in (\ref{ScatAmp}) are 
\begin{equation}
\begin{aligned}
    F_{R} & =z^{-\frac{\alpha'}{4}s} (1-z)^{-\frac{\alpha'}{4}t} K_R(z;k_{i}),
    \\
    F_{L} & = \overline{z}^{+\frac{\alpha'}{4}s} (1-\overline{z})^{+\frac{\alpha'}{4}t} K_L(\overline{z};k_{i})	\,,
\end{aligned}\label{schem_form}
\end{equation}
where $K_R(z;k_{i})$ and $K_L(\overline{z};s,t,u)$ are kinematic factors having the following structure:
\begin{equation}
\begin{aligned}
  K_{R}(z;k_{i}) = \,& P_{1}(k_{i}) +\frac{P_{2}(k_{i})}{z^2}+ \frac{P_{3}(k_{i})}{(1-z)^2} + \frac{Q_{1}(k_{i})}{z} + \frac{Q_{2}(k_{i})}{1-z}  + \frac{Q_{3}(k_{i})}{z(1-z)}\\
&+  \frac{Q_{4}(k_{i})}{z(1-z)^2} + \frac{Q_{5}(k_{i})}{z^2(1-z)} + \frac{z\ Q_{6}(k_{i})}{(1-z)}+ \frac{z\ Q_{7}(k_{i})}{(1-z)^2}   \,,
\end{aligned}\label{Kine1}
\end{equation}
and 
\begin{equation}
\begin{aligned}
  K_{L}(\overline{z};k_{i}) = \,& \overline{P}_{1}(k_{i}) +\frac{\overline{P}_{2}(k_{i})}{\overline{z}^2}+ \frac{\overline{P}_{3}(k_{i})}{(1-\overline{z})^2} + \frac{\overline{Q}_{1}(k_{i})}{\overline{z}} + \frac{\overline{Q}_{2}(k_{i})}{1-\overline{z}}  + \frac{\overline{Q}_{3}(k_{i})}{\overline{z}(1-\overline{z})}\\
&+  \frac{\overline{Q}_{4}(k_{i})}{\overline{z}(1-\overline{z})^2} + \frac{\overline{Q}_{5}(k_{i})}{\overline{z}^2(1-\overline{z})} + \frac{\overline{z}\ \overline{Q}_{6}(k_{i})}{(1-\overline{z})}+ \frac{\overline{z}\ \overline{Q}_{7}(k_{i})}{(1-\overline{z})^2} \,.
\end{aligned}\label{Kine2}
\end{equation}

\noindent Here, $P_{i}(k_{i})$ and $Q_{i}(k_{i})$ are kinematic factors, whose explicit forms are collected in (\ref{PQ}). These are given by polynomials of external momenta $k_{i}$ and do not contain any $s,t$, $u$ poles. Thus, the poles arise only from the integration over moduli variable $z$. 

One can evaluate (\ref{ScatAmp}) applying the following integration formula
\begin{multline}
  \int  d^2 z ~ z^{-x-a_1} (1-z)^{-y-b_1} \overline{z}^{x-a_2} (1-\overline{z})^{y-b_2} \\
   = 2\pi \frac{\Gamma[1-a_1-x] \,\Gamma[1-b_1-y] \Gamma[-1+a_2+b_2-x-y]}
 {\Gamma[a_2-x]\,\Gamma[b_2-y]\Gamma[2-a_1-b_1-x-y]}\,.
\end{multline}
It is useful to write this integral as
\be
  \int  d^2 z ~ z^{-x-a_1} (1-z)^{-y-b_1} \overline{z}^{x-a_2} (1-\overline{z})^{y-b_2}=I(a_1,b_1) \overline{I}(a_2,b_2)	\,,
\ee
where
\be
I(a_1,b_1) = 2 \frac{\Gamma[1-a_1-x] \,\Gamma[1-b_1-y]\,}{\Gamma[2-a_1-b_1-x-y]}, 
\ee
\be
 \overline{I}(a_2,b_2)=\pi \frac{ \Gamma[-1+a_2+b_2-x-y]}
   {\Gamma[a_2-x]\,\Gamma[b_2-y]\,}\,.
\ee
Using the gamma function property, $\Gamma(1+z) = z \Gamma(z)$, one can show that all the gamma functions cancel out. After some straightforward computation, we obtain the four dilaton scattering amplitude in the form
\begin{equation}
\begin{aligned}
  M_4^{\scriptscriptstyle D}(1,2,3,4) = g_c^4 \ \varepsilon^{\scriptscriptstyle (D)}_{1 \mu_1\nu_1} \cdots \varepsilon^{\scriptscriptstyle (D)}_{4 \mu_4\nu_4} A_{R}(k_{i}){}^{\mu_1 \mu_2 \mu_3 \mu_4}A_{L}(k_{i}){}^{\nu_1 \nu_2 \nu_3 \nu_4}\ ,
\end{aligned}
\end{equation}
where
\begin{equation}
\begin{aligned}
  A_{R}(k_{i}) &= \, I(0,0)P_{1}(k_{i}) +I(2,0)P_{2}(k_{i})+ I(0,2)P_{3}(k_{i}) 
  \\
  &+ I(1,0)Q_{1}(k_{i}) +I(0,1) Q_{2}(k_{i})  + I(1,1)Q_{3}(k_{i}) 
  \\
  &+ I(1,2)Q_{4}(k_{i}) +I(2,1) Q_{5}(k_{i})  + I(-1,1)Q_{6}(k_{i}) + I(-1,2)Q_{7}(k_{i}) 
  \,,
\end{aligned}
\end{equation}
and
\begin{equation}
\begin{aligned}
  A_{L}(k_{i}) &= \, \overline{I}(0,0)\overline{P}_{1}(k_{i}) +\overline{I}(2,0)\overline{P}_{2}(k_{i})+ \overline{I}(0,2)\overline{P}_{3}(k_{i})
  \\
  &+ \overline{I}(1,0)\overline{Q}_{1}(k_{i}) +\overline{I}(0,1)\overline{Q}_{2}(k_{i})  + \overline{I}(1,1)\overline{Q}_{3}(k_{i}) 
  \\
  &+ \overline{I}(1,2)\overline{Q}_{4}(k_{i}) +\overline{I}(2,1) \overline{Q}_{5}(k_{i})  + \overline{I}(-1,1)\overline{Q}_{6}(k_{i}) + \overline{I}(-1,2)\overline{Q}_{7}(k_{i}) \,.
\end{aligned}
\end{equation}

To simplify the computation, we introduce a suitable gauge choice for the auxiliary vector $\overline{k}^{\mu}$ introduced in the dilaton polarization tensor \cite{Metsaev:1987zx}. In terms of light-cone kinematics, we set 
\begin{equation}
\quad \overline{k}^{+}=\frac{1}{k^{-}}\,, \quad \overline{k}^{-} = 0\,, \quad \overline{k}^{m} = 0\,, \qquad m = 2,\cdots , D-1\,,
\label{}\end{equation}
We also introduce the Mandelstam variables $s,t$ and $u$ defined as
\begin{equation}
  k_1 \cdot k_{2} = k_{3}\cdot k_{4} = - \frac{s}{2}\,, \quad k_{2} \cdot k_{3} =k_{1}\cdot k_{4} = - \frac{t}{2}\,, \quad  k_{1} \cdot k_{3} = k_{2}\cdot k_{4}= - \frac{u}{2}\,,
\label{}\end{equation}
where $s+t+u=0$ for massless fields. 
The scattering amplitude has to be independent of the gauge choice. After a tedious algebra, we get
\begin{tcolorbox}
\begin{equation}
\begin{aligned}
  M^{\scriptscriptstyle D}_4 = \frac{\alpha'^{-5} g_c^4 C_{S^2} \pi}{2(D-2)^2} {f(s, t, u) \over  s t u (s-4/\alpha') (s+4/\alpha') (t-4/\alpha') (t+4/\alpha') (u-4/\alpha') (u+4/\alpha')} \,, \nonumber
\end{aligned}
\end{equation}
where $f(s,t,u)$ is a polynomial of $s,t$ and $u$ of degree 12, 
\begin{equation}
\begin{aligned}
 & f(s,t,u) \\
 = & - 2^{10}(D-2)^2 (s^2 + t^2 + u^2)^2 
  \\ 
  &+ (\alpha')^{2} \ 2^{6} \Big( (D-2)^2 (s^2+t^2+u^2)^3 - 4\big(3(D-2)(D-14)+32 \big) s^2 t^2 u^2\Big)
   \\
   & -(\alpha')^{4}\Big((D-2)^2 (s^2 + t^2 + u^2)^4 -32 \big((D-2)(D-11)+8\big) s^2 t^2 u^2 (s^2 + t^2 + u^2) \Big)
 \\
 & + (\alpha')^6\ 2  (D-2) s^2 t^2 u^2 (s^2 + t^2 + u^2)^2 
 \\
 & -(\alpha')^8 s^4 t^4 u^4\ .  \label{4pointdilaton}
 \end{aligned}
\end{equation}
\end{tcolorbox}
\noindent
Here, we explicitly displayed the dependence on spacetime dimension,  $D=26$, and included the  factor $C_{S^2}=\frac{8\pi}{\alpha' g_c^2}$ that for simplicity we omitted at the beginning of the calculation. See Appendix B. One can easily note that the location of poles exactly matches with the mass spectrum listed in Table \ref{table1} and that the residue at these poles fits perfectly with product of two three-point amplitudes. We relegate details of this confirmation to Appendix B. 

The scattering amplitude (\ref{4pointdilaton}) is given by `rational function' of kinematic invariants, $s, t, u$. This indicates that string theory over the new vacuum is nothing but a field theory. On the other hand, the scattering amplitude is manifestly invariant under the $s\leftrightarrow t\leftrightarrow u$ channel duality. We see that the string theory description nicely sums over field theory Feynman diagrams over all channels at once. 

The behavior of (\ref{4pointdilaton}) at low- or high-energy regime (relative to the string tension $T = 1/(2 \pi \alpha')$) can be analyzed straightforwardly. They are equivalent to the limits of $\alpha'$ to zero or to infinity, respectively. In the limit $\alpha^{\prime}\rightarrow0$, the amplitude fits perfectly to the four dilatons  amplitude \cite{Pasukonis:2005db} in ordinary string theory
\be
 M^{\scriptscriptstyle D}_{4}=\pi^2g_c^2\frac{(s^2+t^2+u^2)^2}{stu}\ .
\ee
In order to match the result in the reference \cite{Pasukonis:2005db} the relation 
\be
(s^2+t^2+u^2)^2=4(s^2t^2+s^2u^2+t^2u^2)
\ee
must be used with $s+t+u=0$.
From the pole structure of (\ref{4pointdilaton}) when $\alpha^{\prime}\rightarrow0$,  we deduce that there are only massless excitations.  On the other hand, in the tensionless limit, $\alpha'\to \infty$, the leading term of the four-point scattering amplitude behaves as
\begin{equation}
  M^{\scriptscriptstyle D}_{4} \quad \rightarrow \quad -\frac{4\pi^2(g_c\alpha')^2}{(D-2)^2} s  t u\, + \mbox{(sub-leading pole terms)}\, .
\label{amplimit}\end{equation}
From (\ref{amplimit}),  we see that the four-point scattering amplitude is dominated by  contact interactions and grows arbitrarily large, eventually violating the elastic unitarity bound. This is simply an indication that  string theory  quantized over the new vacuum just behaves as a non-renormalizable field theory. We can also extract the field contents in this limit by examining the sub-leading term in $1/\alpha'$. It is straightforward to see that this term is proportional to $1/stu$. This fits perfectly with the spectrum in Table \ref{table1} in the tensionless limit $\alpha' \rightarrow \infty$.


\section{The torus partition function and one-loop cosmological constant}\label{partition_func}
In this section, we study quantum aspects of  string theory quantized over the new vacuum.  In the previous sections, we presented various arguments that the theory is in fact a field theory containing dynamical gravity. One would thus expect that the quantum effects display more field theoretic properties than string theory properties. We will present an evidence for this through explicit {\sl ab initio} computation of the one loop vacuum amplitude. 

\subsection{Field theory toy model}
A feature of  string theory over the new vacuum is that it contains ghost fields, fields of negative norm. To gain further intuition about the string partition function in the presence of ghost fields, we first compute the one-loop partition function in a simple quantum field theory. Specifically, we consider a theory containing a healthy free massless scalar field $\phi_2$ and two ghost scalar fields $\phi_1, \phi_3$ with opposite values of $m^2$. The action for this theory is given by
\be
S=\frac{1}{2}\int\big[ -\phi_1(\square-m^2)\phi_1+\phi_2\square\phi_2-\phi_3(\square+m^2)\phi_3\big]\label{action}\ .
\ee
The one-loop cosmological constant is given by $W=-\log Z$, where $Z$ is the one-loop partition function
\be
Z=\int \!\! D\phi_1 D\phi_2 D\phi_3\,\, \text{e}^{-S}\ .
\ee

The computation of the partition function is tricky since the path integral in $\phi_1$ and $\phi_3$ is ill-defined at its disposal. We will deform the contour and adopt the following prescription \cite{Gibbons:1978ac}
\be
\phi_1\rightarrow\text{i}\ \phi_1\quad;\quad \phi_3\rightarrow\text{i}\ \phi_3\ .
\ee
After the deformation, the path integral is convergent, but we get an extra imaginary factor. We adopt the normal ordering scheme of dropping the (imaginary) infinite normalization factor, and express the one-loop cosmological constant as 
\be
W=\int\frac{d^Dk}{(2\pi)^D}\int_0^\infty\frac{ds}{s}e^{-\frac{1}{2}k^2}\Big( e^{-m^2s}+ 1+ e^{+m^2s}\Big)\label{qftresult}\ .
\ee

\noindent This toy model illustrates the point that even in the presence of ghost fields, with suitable analytic continuation of the path integral, equivalently, of Feynman rules, the one-loop cosmological constant and partition function can be computed. This toy model will also serve for establishing parallels with string theory computations below. 

\subsection{One-loop partition function with the unconventional vacuum}
The full partition function including the $b$, $c$ worldsheet ghosts can be computed using the conventional definition of partition function. In the operator formalism, the one-loop partition function over a torus of complex structure $\tau$ is given by
\be
\text{Z}(\tau)=\text{Tr}\big[(-1)^F(-1)^{\overline{F}}c_0b_0\overline{c}_0\overline{b}_0\text{exp}\big(2\pi\text{i}\tau_1P-2\pi\tau_2H \big)\big]\label{partition}
\ee
\be
P=L^{\rm tot}_0-\bar{L}^{\rm tot}_0 \, ,  \qquad \qquad \qquad H=L^{\rm tot}_0+\bar{L}^{\rm tot}_0 \,.
\ee
Here, the  $F$ and $\overline{F}$ are fermionic number operators acting on the right-moving and the left-moving parts, respectively. The $L^{\rm tot}_0$ and  $\bar{L}^{\rm tot}_0$ are the zero-mode Virasoro operators defined earlier in (\ref{virasoro_bc_0}). We shall rewrite these operators more compactly, making the distinction between the bosonic and the ghost level operators,
\bea \nonumber
L^{\rm tot}_0      & = & \frac{1}{2}\alpha_0^2+{\bf N}_B+{\bf N}_g-1 \\ \nonumber
\bar{L}^{\rm tot}_0 & = & \frac{1}{2}\overline{\alpha}_0^2+ \overline{{\bf N}}_B+\overline{{\bf N}}_g+1\ .
\eea

Substituting the Virasoro operators into the expression (\ref{partition}), after some algebra, we recast the partition function as 
\bea
\text{Z}(\tau) & = & \text{e}^{-4\pi\text{i}\tau_1}\\\nonumber
{} & {} &\text{Tr}_0\big[\text{e}^{+\pi\tau \alpha_0^2}\big]\text{Tr}_R\big[\text{e}^{+2\pi\text{i}\tau{\bf N}_B}\big]\text{Tr}_R\big[(-1)^{F}c_0b_0\text{e}^{+2\pi\text{i}\tau{\bf N}_g}\big]\\ \nonumber
{} & {} &\text{Tr}_0\big[\text{e}^{-\pi\bar{\tau}\overline{\alpha}_0^2}\big] \text{Tr}_L\big[\text{e}^{-2\pi\text{i}\overline{\tau}\overline{\bf N}_B}\big]\text{Tr}_L\big[(-1)^{\overline{F}}\overline{c}_0\overline{b}_0\text{e}^{-2\pi\text{i}\overline{\tau}\overline{\bf N}_g}\big]\, . \nonumber
\eea
We do not present the computation for the contribution of right-moving oscillators since it is the same as usual string theory. It is given by
\be
\text{Tr}_R\big[\text{e}^{2\pi\text{i}\tau{\bf N}_B}\big]\text{Tr}_R\big[(-1)^{F}c_0b_0\text{e}^{2\pi\text{i}\tau{\bf N}_g}\big]=\Big[\sum_{N=0}^{\infty}P\big(N\big)\Big(\text{e}^{2\pi\text{i}\tau}\Big)^{N}\Big]^{(D-2)}\ . 
\ee
Here, $P(N)$ is the number of partitions of the level. We have used  this unconventional expression of the partition function, as it will be useful to perform the integration over the moduli space of the torus.  

We now present a collection of some useful information needed for the computation of the contribution of the left-moving oscillators to the partition function. The following relations hold for the left-moving sector, where we use the definition in footnote 6,
\bea\label{useful}
\overline{\bf N}_B| \overline{N}_{Bn}\rangle & = & -n\overline{N}_{Bn} | \overline{N}_{Bn}\rangle \\ \nonumber
\langle\overline{N}_{Bn}| \overline{N}_{Bn}\rangle & = &  (-1)^{\overline{N}_{Bn}}\\ \nonumber
\langle\overline{N}_{Bn}|\overline{\bf N}_B| \overline{N}_{Bn}\rangle & = & -n\overline{N}_{Bn} (-1)^{\overline{N}_{Bn}}\,.\nonumber
\eea
We first compute the matter oscillator contribution. By definition, the trace is given by
\bea\label{partition_left}
\text{Tr}_L\big[\text{e}^{-2\pi\text{i}\overline{\tau}\overline{\bf N}_B}\big] & = & \Big[\prod_{n=1}^{\infty}\sum_{\overline{N}_n=0}^{\infty}\frac{\langle\overline{N}_{Bn}|
\text{e}^{-2\pi\text{i}\overline{\tau}\overline{\bf N}_B}|\overline{N}_{Bn}\rangle}{\langle\overline{N}_{Bn}|\overline{N}_{Bn}\rangle}\Big]^D \\ \nonumber
{} & = & \Big[ \prod_{n=1}^{\infty}\sum_{\overline{N}_{Bn}=0}^{\infty}\text{e}^{2\pi\text{i}\overline{\tau}n\overline{N}_{Bn}}\Big]^D\,,  \nonumber
\eea
where the normalization in the denominator is taken into the definition of the trace, because the states $|\overline{N}_{Bn}\rangle$ are not ortho-normal. To obtain the expression in the second line of (\ref{partition_left}) we used the last relation of (\ref{useful}) .

The sum in the second line of (\ref{partition_left}) can be performed only after regularization,  as the  geometric series does not converge because of the fact that $| \text{e}^{2\pi\text{i}\overline{\tau}n}|>1$. Here, we proceed by analytically extending the sum of the geometric series to an arbitrary value of the parameter, viz. 
\be
\sum_{n=0}^{\infty}q^n=\frac{1}{1-q}\,, \qquad \forall \ q \ .
\ee
After the regularization, we get
\be
\text{Tr}_L\big[\text{e}^{-2\pi\text{i}\overline{\tau}\overline{\bf N}_B}\big]  =\Big[ \prod_{n=1}^{\infty}\frac{1}{1-\text{e}^{2\pi\text{i}\overline{\tau}n}}\Big]^D\ .
\ee
Note that this is a formal expression even after regularization: the infinite product does not converge. However, we can proceed in the other way around by  performing the infinite product first in the second line of (\ref{partition_left}). By doing so, we can assign a series to this expression 
\be
\text{Tr}_L\big[\text{e}^{-2\pi\text{i}\overline{\tau}\overline{\bf N}_B}\big]  = \Big[\sum_{\overline{N}=0}^{\infty}P\big(\overline{N}\big)\Big(\text{e}^{2\pi\text{i}\overline{\tau}}\Big)^{\overline{N}}\Big]^D\ .
\ee
This follows from Euler generating function that states that 
\be
(1 + q + q^2 + q^3 + \cdots)(1 + q^2 + q^4 + q^6 + \cdots)(1 + q^3 + q^6 + q^9 + \cdots)\cdots=\sum_{n=1}^{\infty}P(n)q^n\label{euler}\ ,
\ee
which holds for $| q| <1$. Here, we extend (\ref{euler}) to arbitrary $q$ as a regularization prescription.

We next compute the contribution of left-moving sector of $b$, $c$ ghost fields to the partition function:
\bea
\text{Tr}_L\big[(-1)^{\overline{F}}\overline{c}_0\overline{b}_0\text{e}^{-2\pi\text{i}\overline{\tau}\overline{\bf N}_g}\big] & = & \langle \downarrow  | \overline{c}_0\overline{b}_0| \uparrow \rangle\Big[\prod_{n=1}^{\infty}\sum_{\overline{N}_{gn}=0}^{1}\langle\overline{N}_{gn}|(-1)^{\hat{\overline{F}}}\text{e}^{-2\pi\text{i}\overline{\tau}\overline{\bf N}_g}| \overline{N}_{gn}\rangle\Big]^2\\ \nonumber
{} & = &  \langle \downarrow  | \overline{c}_0|  \downarrow \rangle(\langle 0_g |  0_g \rangle)^2\Big[\prod_{n=1}^{\infty}(1-\text{e}^{2\pi\text{i}\overline{\tau}n})\Big]^2\ .
\eea
Again, we should treat the last expression as a formal expression, since this infinite product as it is does not converge.

Identifying $X_L$ and $X_R$ as in (\ref{xlxriden}) and  collecting all the pieces,  we get the final expression of the partition function
\begin{tcolorbox}
\bea\nonumber
\!\!\text{Z}(\tau) \! & = & \!\! \int\frac{d^Dk}{(2\pi)^D}e^{-\pi\tau_2 \alpha^{\prime}k^2} \text{e}^{-4\pi\text{i}\tau_1} \Big[\sum_{N=0}^{\infty}P\big(N\big)\Big(\text{e}^{2\pi\text{i}\tau}\Big)^{N}\Big]^{(D-2)} \Big[\sum_{\overline{N}=0}^{\infty}P\big(\overline{N}\big)\Big(\text{e}^{2\pi\text{i}\overline{\tau}}\Big)^{\overline{N}}\Big]^{(D-2)} .
\eea
\end{tcolorbox}
In conventional string theory, for studying  modular invariance,  one rewrites the partition function in terms of the Dedekind function. The point here is that it is impossible to do the same in the quantization of  string theory over the new vacuum. The contribution of left-moving sector does not converge in the lower-half complex plane, $\overline{\tau} = \tau_1 - i\tau_2$ for $\tau_2 > 0$. It is well-known the Dedekind function does not admit any analytical  continuation. For this reason, we conclude that this partition function is not modular invariant.

As the partition function is not modular invariant, instead of performing the integration over the moduli space of  the torus in the fundamental domain, we now need to perform the integration over the full strip $\tau_2>0$ and $| \tau_1 |<\frac{1}{2}$. Moreover, we can perform first the integration in the $\tau_1$ direction. Interestingly enough, after the integration in $\tau_1$, which is equivalent to imposing the level-matching  condition, we  get a finite contribution which is in perfect agreement with the finite number of degrees of freedom.  This result is expected from the simple quantum field theory model (\ref{qftresult}),
\begin{tcolorbox}
\vskip-0.3cm
\bea\nonumber
\text{Z}(\tau_2) := \int_{-\frac{1}{2}}^{\frac{1}{2}}\text{Z}(\tau)d\tau_1 
&=& \int\frac{d^Dk}{(2\pi)^D}e^{-\pi\tau_2 \alpha^{\prime}k^2} 
\Big[ (D-2)^2 \nonumber \\
&+ &\frac{1}{2}(D-2)(D+1)\ \text{e}^{+4\pi \tau_2}
         +  \frac{1}{2}(D-2)(D+1)\  \text{e}^{-4\pi \tau_2}\Big] \ .
         \nonumber
\eea
\end{tcolorbox}
\noindent
Indeed, inside the bracket, the first term is the contribution of massless string gravity (metric, Kalb-Ramond, dilaton fields), the second term is the contribution of Fierz-Pauli massive spin-two field of mass-squared $-4/\alpha'$, while the last term is the contribution of Fierz-Pauli massive spin-two field of mass-squared $+4/\alpha'$.


\section{Conclusions and outlooks}\label{conclusions}

In this work, we studied {\sl ab initio} quantization of closed bosonic string over the new vacuum (\ref{vc1})  within the operator formalism. The choice of the new vacuum (\ref{vc1})  led to a string theory with a  finite number of degrees of freedom. Specifically, this construction provided a novel string reformulation of various seemingly disparate field theories of string gravity and a pair of spin-two Pauli-Fierz fields. Feynman diagrams in these field theories were computed efficiently by the string worldsheet moduli integration. In order to obtain a well-defined worldsheet correlator, it was necessary to adopt backward time ordering (\ref{torder}) for the left-moving sector. We showed that this choice  is compatible with the normal ordering for the left-moving oscillators. We also clarified the origin of negative norm for the new vacuum, an important point which was overlooked in the previous works.  

We also developed in section \ref{amplitudes} the generating function for the tree-level scattering amplitudes (\ref{fullA}).  It is worth to emphasize that it was computed in the operator formalism over the new vacuum (\ref{vc1}). As a specific but nontrivial case, we computed the four-point dilaton scattering amplitude. This particular contribution involved the traceless part of the polarization of the massless spin-2 state (\ref{trace_part}). Notice that if we were to use the KLT relations together with the metric sign flipping HSY prescription, we would inevitably encounter an ambiguity in the identification of the spacetime metric in the dilaton polarization (\ref{trace_part}). As the only allowed contractions within the HSY prescription are  $\eta_{\mu\nu}\zeta^{\mu}\zeta^{\nu}$ and $(-\eta_{\mu\nu})\overline{\zeta}^{\mu}\overline{\zeta}^{\nu}$,  this prescription did not hold for  $\pm\eta_{\mu\nu}\zeta^{\mu}\overline{\zeta}^{\nu}$. In our {\sl ab initio} formulation of quantized string theory, no such ambiguity arises.  The spectrum matched with the field contents of HSY \cite{Huang:2016bdd}. There were three spin-2 fields:  two massive ghost  field and a massless state containing the metric, the Kalb-Ramon field and the dilaton. 

In the previous  work \cite{Hwang:1998gs}, a similar study was performed within the BRST quantization. Because of the existence of indefinite metric states in the BRST invariant sector, the work \cite{Hwang:1998gs} concluded that the resulting theory is inconsistent. In particular, the massless spin-two field is a ghost. Unlike \cite{Hwang:1998gs}, however, here we choose the vacuum to be a negative norm state. As a consequence of this choice, the massless spin-two state becomes a healthy, unitary field. Of course, the theory still contains two massive negative norm  spin-2 fields. 

 Perhaps, one of the most intriguing results of this work is the one-loop partition function. Despite the presence of negative norm states and negative energy levels (\ref{useful}) for the left-moving sector, the partition function can be regularized. For the right-moving sector, we obtained the usual contribution. Left sector, by itself, was ill-defined. All the series and infinite products involved in the calculation did not converge.  We treated them as formal expressions depending on the moduli of the torus $(\tau_1,\tau_2)$. As expected, the partition function $Z(\tau_1,\tau_2)$ is not modular invariant (something similar happens in the tensionless limit \cite{Adamo:2013tsa, Yu:2017bpw}). We associated the lack of modular invariance with the presence of negative norm states. After integration over $\tau_1$, a finite number of terms remained in the partition function. Interestingly   the partition function properly makes  the counting of the degrees of freedom of the theory, and it coincides with the QFT result. Remarkably  this QFT result has a stringy  origin and we obtained it passing through not so well-defined mathematical expressions. 

One of the open questions is whether the tensionless limit commutes  with the quantization. Our work provides a platform  for answering this question. It would be interesting to explore the tensionless limit within this scheme. In Appendix \ref{tensionlesslimitapp}, we present how to take the limit from the tensile mode expansion to get the tensionless one. One may look at the amplitude (\ref{4pointdilaton}) and consider the $\alpha^{\prime}$ infinite limit. It is straightforward to see that the amplitude reduces to (\ref{amplimit}).
Notice that no negative-norm particle propagates as intermediate states after such limit. 

We can go beyond the tree-level amplitudes. Using the mode expansion on the torus, one can compute  the loop amplitudes for finite $\alpha^{\prime}$ within the scheme presented here.  We expect  the tensionless limit of such amplitudes provide the ambitwistor one loop amplitudes. We also believe  that, proceeding in this way, the moduli integration problem could be overcome. 

One of the most interesting future directions is the supersymmetric generalization. As argued in \cite{Huang:2016bdd,Leite:2016fno,Azevedo:2017yjy}, Type II and heterotic superstring would be safe from the ghost degrees of freedom. It would be interesting to examine the unitarity of scattering amplitudes and extension of no-ghost theorem for the alternative vacuum choice for tensile string \cite{forthcoming}. 
For the conventional vacuum, we know that the supersymmetric partition function vanishes due to the cancelation between bosonic and fermionic contributions and that the torus partition function exhibits the modular invariance.

The tree-level scattering amplitude for ambitwistor string and type II HSY string are equivalent to the type II supergravity amplitude\cite{Mason:2013sva,Huang:2016bdd,Leite:2016fno}. Apparently, the supersymmetric scattering amplitudes for the unconventional vacuum choice are insensitive with respect to $\alpha'$ at tree-level. As we can go beyond the tree-level amplitudes,  further supersymmetric generalization of our quantization scheme should be useful to study the validity of this mystery at higher genus. 

\section*{Acknowledgements}
We are 
 to Gleb Arutyunov, Junegone Chay, Harald Grosse, Daniel Grumiller, Taejin Lee, Jeong-Hyuck Park, Sasha Polyakov, Massimo Porrati and Jennie Traschen for helpful discussions, and to Fidel  Schaposnik Massolo for the help with the Mathematica code. We thank hospitality of University of Massachusetts-Amherst, New York University, Princeton University and Technical University of Vienna, University of Chicago, UC-San Diego during the course of this work. 
Preliminary results of this work were reported at "String and M-Theory Geometries" workshop (January 22 - 27, 2017, Banff, Canada), "Recent Advances in T/U-dualities and Generalized Geometries" workshop (June 6-9, Zagreb, Croatia), and ESI Program "Quantum Physics and Gravity" (May 29 - July 13, Vienna, Austria). We acknowledge useful comments from participants at these workshops.

\appendix


\section{Tensionless Limit}\label{tensionlesslimitapp}


To see the connection between our new vacuum (\ref{newvac}) and the ambitwistor vacuum defined for tensionless and null string, we look at the mode expansion of the tensile and tensionless string. Although these two mode expansions look very different, the latter can be obtain from the former by the rescaling, 
\bea
\tau\rightarrow \epsilon \tau, \quad T \rightarrow \epsilon T_o \quad \mbox{for} \quad \epsilon \rightarrow  0 \quad \mbox{while} \quad
Y^\mu = \sqrt{T_0} X^\mu = \mbox{finite}
 \ , 
\eea
where $T_0$ is an arbitrary reference scale of the tension. 
The new coordinate $Y^\mu$ has spacetime scaling weight 0. Such rescaling is motivated by the expectation that the tensionless limit restores spacetime conformal invariance, much like the massless limit of point particle does. 
This is equivalent to the ultrarelativistic limit studied originally in \cite{Lizzi:1994rn} and more recently in \cite{Bagchi:2009my,Bagchi:2009pe,Bagchi:2013bga}. 

First, we demonstrate that, starting from mode expansion of tensile string, we can recover the mode expansion of tensionless or null string. Following \cite{Gamboa:1989px}, \cite{Gamboa:1989zc}, the mode expansion for tensionless string or null string can be written as
\be
Y^\mu(\tau, \sigma)=\big(\frac{1}{2\sqrt{\pi}}\sum_{n=-\infty}^{\infty}Y_n^{\mu}\text{e}^{\text{i}n\sigma}\big)+\big(\frac{1}{2\sqrt{\pi}}\sum_{n=-\infty}^{\infty}P_n^{\mu}\text{e}^{\text{i}n\sigma}\big)\tau\, ,\label{modeexpaTzero}
\ee
while for tensile string,
\be
X^\mu(\tau, \sigma) =
X_0^\mu + \frac{\alpha'}{2}  P^\mu \tau  +  i \sqrt{\alpha' \over 2} \sum_{n \ne 0} {1 \over n} \overline \alpha^\mu_n  
e^{ -  i n (\tau + \sigma)}+ i \sqrt{\alpha' \over 2} \sum_{n \ne 0} {1 \over n} \alpha^\mu_n  
e^{ -  i n (\tau - \sigma)}\ .
\ee
Rescaling the string tension $T$, worldsheet time $\tau$ and the string coordinates $X^\mu$ as specified above, we can write the mode expansion as
\be
Y^\mu(\tau, \sigma) =
Y_0^\mu + \frac{1}{4\pi}P^\mu \tau  + { i \over 2\sqrt{\pi}} \sum_{n \ne 0} {1 \over n} \overline \alpha^\mu_n  
e^{ -  i n (\epsilon \tau + \sigma)}+{ i \over 2\sqrt{\pi}}  \sum_{n \ne 0} {1 \over n} \alpha^\mu_n  
e^{ -  i n (\epsilon \tau - \sigma)}\ .
\label{mode1111}\ee
Taking into account (\ref{C_A_op}), the mode expansion can be separated into two pieces, one that depends on $X$'s and one depending only on $P$'s. When $T\rightarrow0$ the previous mode expansion can be expanded as \cite{Bagchi:2015nca,Bagchi:2016yyf}
\bea
Y^\mu(\tau, \sigma)  & = & Y_0^{\mu}+ \frac{1}{2\sqrt{\pi}}  \sum_{n \ne 0}Y_n^ {\mu}\ \text{e}^{\text{i}n\sigma}+\frac{1}{2\sqrt{\pi}}  \sum_{n \ne 0}Y_{-n}^ {\mu}\ \text{e}^{-\text{i}n\sigma}\\ \nonumber
{} & + & \frac{1}{\pi}P^{\mu}\tau+\frac{\text{i}}{2\sqrt{\pi}}  \sum_{n \ne 0}\frac{1}{2n\epsilon}(1-\text{i}n\epsilon\tau)P_n^{\mu}\ \text{e}^{\text{i}n\sigma}
+\frac{\text{i}}{2\sqrt{\pi}}  \sum_{n \ne 0}\frac{1}{2n\epsilon}(1-\text{i}n\epsilon\tau)P_{-n}^{\mu}\ \text{e}^{-\text{i}n\sigma} .
\eea
We see that the divergent part in the second line cancels out each other and that, after rescaling the zero modes $Y_0^\mu$'s and  $P^\mu$'s, we get the mode expansion (\ref{modeexpaTzero}).  Note that the mode expansion (\ref{mode1111}) is the standard one, and it is obtained in the conformal gauge. In particular, no singular gauge choice for the worldsheet metric is required. 

With the above rescaling, we obtain the ambitwistor vacuum in the tensionless limit 
\bea  \nonumber
Y_n^{\mu}\vert 0\rangle & = & 0\ , \\ 
P_n^{\mu}\vert 0\rangle & = & 0\,, \qquad \text{for}~n>0\,,
\eea
from the new vacuum prescription (\ref{vc1}). Notice that at the level of the mode expansion, this limit is regular. Therefore, one would expect ambitwistor string would arise from the tensionless limit of quantized string over the new vacuum (\ref{vc1}).

For completeness, we compute the tensionless limit of (\ref{two_point_ft}). We take into account (\ref{ztrans}) and take the tensionless limit as above. After taking the tensionless limit and dropping  a linear divergent term, which appears after taking the limit $\epsilon \rightarrow 0$,   we get
\be
  \big\langle 0\big|T\big[X^{\mu}(z_{i},\overline{z}_{i}) X^{\nu}(z_{j},\overline{z}_{j})\big]\big|0\big\rangle =\frac{i}{4\pi}(\tau_i-\tau_j)\frac{z_i+z_j}{z_i-z_j}\ ,
\ee
where after the limit we rename the $z$ variable as 
\be
z = \exp( i\sigma)\ .
\ee
A similar result was obtained  for the null string theory \cite{Casali:2017zkz}. 

As a consequence of the vacuum choice the Virasoro generators satisfy the algebra
\bea \nonumber\label{viraalgebra}
\big[L_m, L_n\big ] & = & (m-n)L_{m+n}+\frac{c}{12}m(m^2-1)\delta_{m+n,0}\\
\big[\overline{L}_m, \overline{L}_n\big ] & = & (m-n)\overline{L}_{m+n}-\frac{\bar c}{12}m(m^2-1)\delta_{m+n,0}\, .
\eea
Note that the central charge of the right-moving sector is positive but the left-moving sector is negative, $c= \bar c=D-2$.
At this point, it is useful to combine the two sets of Virasoro generators into 
\bea\nonumber\label{redefvira}
{\cal L}_m \ & = & \, (L_m-\overline{L}_{-m}) \\ 
{\cal M}_m & = & \epsilon(L_m+\overline{L}_{-m})
\eea
where we introduced a parameter $\epsilon$ that will be related eventually to the tension of the string. The above Virasoro algebra now reads
\bea \nonumber\label{viraalgebraLM}
\big[\ {\cal L}_m, \ {\cal L}_n \ \big ] & = & (m-n){\cal L}_{m+n}+\frac{c +\bar c}{12}m(m^2-1)\delta_{m+n,0}\\ \nonumber
\big[{\cal M}_m, {\cal M}_n\big ] & = & \epsilon^2 \Big[(m-n){\cal L}_{m+n}+\frac{c +\bar c}{12}m(m^2-1)\delta_{m+n,0}\Big]\\ 
\big[ {\cal L}_m, \ {\cal M}_n\big ] & = & (m-n) {\cal M}_{m+n}\, .
\eea
As the tow copies of the Virasoro algebra (\ref{viraalgebra}) have opposite central  charges, the algbra (\ref{viraalgebraLM}) contains no central extension\cite{Casali:2017zkz}.  
We see that, in the tensionless limit $\epsilon \rightarrow 0$, we recover the Galilean conformal algebra \cite{Bagchi:2013bga}.

In terms of the generators (\ref{redefvira}),  the physical condition and the mass shell condition (\ref{phy0}) can be recast as \cite{Bagchi:2013bga}
\bea
\langle \text{phys}\mid {\cal L}_0\mid \text{phys}\rangle \ & = & \Delta \langle \text{phys}\mid \text{phys}\rangle \\ \nonumber
\langle \text{phys}\mid {\cal M}_0\mid \text{phys}\rangle & = & \hskip0.1cm \xi  \, \langle \text{phys}\mid \text{phys}\rangle\
\eea
where
\bea
\Delta&=&-\frac{\alpha'}{2} M^2 + N - \overline{N} = 0\,,\\ \nonumber
\xi&=& N+\overline{N}-2 = 0\,.
\eea
In \cite{Bagchi:2013bga}, $\Delta$ was identified with the scaling dimension of the vertex operator while $\xi$ was named `rapidity'. From our treatment, it is clear that they are nothing but the mass-shell condition and the level matching constraint once the physical state condition (\ref{phy0}) is imposed.


\section{Three-point scattering amplitudes}

In this appendix, we recapitulate the details of three-point scattering amplitudes that involve two dilaton fields. These are the building blocks for the factorization of four-point dilaton scattering amplitude we discussed in the text. Using the three-point amplitude, we compute the four-dilaton scattering amplitude (\ref{4pointdilaton})

In general, four-point scattering amplitude is represented by the three-point amplitudes \cite{Leite:2016fno}
\begin{equation}
  M_4(1,2,3,4) = \sum_{i} M_3(1,2,-i) \frac{1}{K_i} M_3(i,3,4) \,,
\label{C1}\end{equation}
where $i$ stands for the all possible intermediate states, and $\frac{1}{K_i}$ is the propagator of the intermediate states. 

For the four-dilaton scattering amplitude (\ref{4pointdilaton}), we shall consider the following three-point amplitudes:
\begin{itemize}
  \item (dilaton)-(dilaton)-(massless string gravity)
\begin{equation}
  M_{\scriptscriptstyle DD m_0} (k_1,\varepsilon^{\scriptscriptstyle (D)}_1;k_2,\varepsilon^{\scriptscriptstyle (D)}_2;k_3,\varepsilon^{}_3) = -g^3_c\Big(\frac{\alpha'}{2}\Big)^{-3} \varepsilon^{\scriptscriptstyle (D)}_{1 \mu_1\nu_1}\varepsilon^{\scriptscriptstyle (D)}_{2 \mu_2\nu_2}\varepsilon^{}_{3 \mu_3\nu_3} T_{R}{}^{\mu_1\mu_2\mu_3} T_{L}{}^{\nu_1\nu_2\nu_3}
\label{3pt_massless}\end{equation}
where $\varepsilon_{\mu_3\nu_3}$ is the polarization tensor for massless spin-2 fields and decomposed as
\begin{equation}
  \varepsilon_{3 \mu_3\nu_3} = \zeta_{3\mu_3} \overline{\zeta}_{3\nu_3}
\label{}\end{equation}
and 
\begin{equation}
\begin{aligned}
  T_{R}{}^{\mu_1\mu_2\mu_3} &= \Big(\frac{\alpha'}{2}\Big)^{2} \Big(\eta^{\mu_1\mu_2} k_1^{\mu_3}+\eta^{\mu_1\mu_3} k_3^{\mu_2}+\eta^{\mu_2\mu_3} k_2^{\mu_1}+\frac{\alpha'}{2} k^{\mu_3}_1 k^{\mu_1}_2 k^{\mu_2}_3\Big)\,,
  \\
  T_{L}{}^{\mu_1\mu_2\mu_3} &= \Big(\frac{\alpha'}{2}\Big)^{2} \Big(\eta^{\mu_1\mu_2} k_1^{\mu_3}+\eta^{\mu_1\mu_3} k_3^{\mu_2}+\eta^{\mu_2\mu_3} k_2^{\mu_1}-\frac{\alpha'}{2} k^{\mu_3}_1 k^{\mu_1}_2 k^{\mu_2}_3\Big)\,.
\end{aligned}\label{}
\end{equation}
\item (dilaton)-(dilaton)-(massive $a_{\mu \nu})$
\begin{equation}
  M_{\scriptscriptstyle DDa} (k_1,\varepsilon_1;k_2,\varepsilon_2;k_3,E_3) = -g^3_c\Big(\frac{\alpha'}{2}\Big)^{-3} \varepsilon^{\scriptscriptstyle (D)}_{1 \mu_1\nu_1}\varepsilon^{\scriptscriptstyle (D)}_{2 \mu_2\nu_2} E^{}_{3 \mu_3\mu_4} S_{R}{}^{\mu_1\mu_2\mu_3\mu_4} S_{L}{}^{\nu_1\nu_2}
\label{3pt_a}\end{equation}
where $E_{\mu\nu}$ is the polarization tensor for the ghost fields with  $m^{2} = 4/\alpha'$ and 
\begin{equation}
\begin{aligned}
  S_{R}{}^{\mu_1\mu_2\mu_3\mu_4} &= \Big(\frac{\alpha'}{2}\Big)^{2}\Big(\eta^{\mu_1\mu_3}\eta^{\mu_2\mu_4} + \eta^{\mu_1\mu_4}\eta^{\mu_2\mu_3} + \frac{\alpha'}{2} \big(\eta^{\mu_2 \mu_4} k_2^{\mu_3} k_3^{\mu_1} + \eta^{\mu_2 \mu_3} k_2^{\mu_4} k_3^{\mu_1} 
  \\
  &\qquad\qquad\quad  - \eta^{\mu_1 \mu_4} k_2^{\mu_3} k_3^{\mu_2} - \eta^{\mu_1 \mu_3} k_2^{\mu_4} k_3^{\mu_2} + \eta^{\mu_1 \mu_2} k_2^{\mu_3} k_2^{\mu_4}\big)
   \\
  &\qquad\qquad\quad - \Big(\frac{\alpha'}{2}\Big)^{2} k_{3}^{\mu_{1}} k_{3}^{\mu_{2}} k_{2}^{\mu_3} k_2^{\mu_4}\Big)
  \\
  S_{L}{}^{\nu_1\nu_2} &= - \frac{\alpha'}{2}\Big(\eta^{\nu_1\nu_2} + \frac{\alpha'}{2}k_1^{\nu_2}k_2^{\nu_1}\Big)\,.
\end{aligned}\label{}
\end{equation}
  \item (dilaton)-(dilaton)-(massive $\overline{a}_{\mu \nu}$)
\begin{equation}
    M_{\scriptscriptstyle DD\overline{a}} (k_1,\varepsilon_1;k_2,\varepsilon_2;k_3,\overline{E}_3) = -g^3_c\Big(\frac{\alpha'}{2}\Big)^{-3} \varepsilon^{\scriptscriptstyle (D)}_{1 \mu_1\nu_1} \varepsilon^{\scriptscriptstyle (D)}_{2 \mu_2\nu_2} \overline{E}_{3 \nu_3\nu_4} \overline{S}_{R}{}^{\mu_1\mu_2} \overline{S}_{L}{}^{\nu_1\nu_2\nu_3\nu_4}
\label{3pt_ba}\end{equation}
where $\overline{E}_{\mu\nu}$ is the polarization tensor for the ghost fields with $m^{2} = -4/\alpha'$ and 
\begin{equation}
\begin{aligned}
  \overline{S}_{R}{}^{\mu_1\mu_2} 	&= \frac{\alpha'}{2}\Big(\eta^{\mu_1\mu_2} - \frac{\alpha'}{2}k_1^{\mu_2}k_2^{\mu_1}\Big)
  \\
  \overline{S}_{L}{}^{\nu_1\nu_2\nu_3\nu_4} &= \Big(\frac{\alpha'}{2}\Big)^{2}\Big(\eta^{\nu_1\nu_3}\eta^{\nu_2\nu_4} + \eta^{\nu_1\nu_4}\eta^{\nu_2\nu_3} -\frac{\alpha'}{2} \big(\eta^{\nu_2 \nu_4} k_2^{\nu_3} k_3^{\nu_1} + \eta^{\nu_2 \nu_3} k_2^{\nu_4} k_3^{\nu_1}
  \\
  &\qquad\qquad\quad - \eta^{\nu_1 \nu_4} k_2^{\nu_3} k_3^{\nu_2} - \eta^{\nu_1 \nu_3} k_2^{\nu_4} k_3^{\nu_2} + \eta^{\nu_1 \nu_2} k_2^{\nu_3} k_2^{\nu_4}\big)
   \\
  &\qquad\qquad\quad - \Big(\frac{\alpha'}{2}\Big)^{2} k_{3}^{\nu_{1}} k_{3}^{\nu_{2}} k_{2}^{\nu_3} k_2^{\nu_4}\Big)\,.
\end{aligned}\label{}
\end{equation}
\end{itemize}

The four-dilaton scattering amplitude consists of three different sectors with respect to the mass of intermediate states:
\begin{equation}
\begin{aligned}
  M^{\scriptscriptstyle D}_4(1,2,3,4) = M_{0}(1,2,3,4) + M_{\frac{4}{\alpha'}}(1,2,3,4) + M_{-\frac{4}{\alpha'}}(1,2,3,4)\,.
\end{aligned}\label{}
\end{equation}
Here, the subscripts on the right-hand side indicate the mass of the intermediate states. If we focus on the $s$-channel, we have
\begin{equation}
\begin{aligned}
  M_{0} &= \sum_{\varepsilon} \frac{M_{DD m_0}(k_1,\varepsilon^{\scriptscriptstyle (D)}_1;k_2,\varepsilon^{\scriptscriptstyle (D)}_2;-k,\varepsilon) M_{DDm_0}(k,\varepsilon;k_3,\varepsilon^{\scriptscriptstyle (D)}_3;k_4,\varepsilon^{\scriptscriptstyle (D)}_4)}{-k^{2}}\,,
  \\
  M_{+\frac{4}{\alpha'}} &= \sum_{E} \frac{M_{DD a}(k_1,\varepsilon^{\scriptscriptstyle (D)}_1;k_2,\varepsilon^{\scriptscriptstyle (D)}_2;-k,E) M_{DDa}(k,E;k_3,\varepsilon^{\scriptscriptstyle (D)}_3;k_4,\varepsilon^{\scriptscriptstyle (D)}_4)}{-k^{2}-\frac{4}{\alpha'}}\,,
  \\
  M_{-\frac{4}{\alpha'}} &= \sum_{\overline{E}} \frac{M_{DD \overline{a}}(k_1,\varepsilon^{\scriptscriptstyle (D)}_1;k_2,\varepsilon^{\scriptscriptstyle (D)}_2;-k,\overline{E}) M_{DD\overline{a}}(k,E;k_3,\varepsilon^{\scriptscriptstyle (D)}_3;k_4,\varepsilon^{\scriptscriptstyle (D)}_4)}{-k^{2}+\frac{4}{\alpha'}}\,,
\end{aligned}\label{3pt^2}
\end{equation}
where $k$ is the intermediate momentum which is given by $k=k_{1}+k_{2} = -k_3 - k_4$. 

In order to evaluate (\ref{3pt^2}), we substitute the above three-point amplitudes in (\ref{3pt_massless}),(\ref{3pt_a}) and (\ref{3pt_ba}), and apply the following completeness relations of the polarization tensors
\begin{equation}
\begin{aligned}
  &\sum_{\zeta} \zeta_{\mu} \zeta_{\nu} =  \sum_{\overline{\zeta}} \overline{\zeta}_{\mu} \overline{\zeta}_{\nu} = \eta_{\mu\nu}\,,
  \\
  & \sum_{E} E_{\mu_{1}\nu_{1}} E_{\mu_{2}\nu_{2}} = a (P^{+}{}_{\mu_{1}\mu_{2}} P^{+}{}_{\nu_{1}\nu_{2}} + P^{+}{}_{\mu_{1}\nu_{2}} P^{+}{}_{\nu_{1}\mu_{2}} - \frac{2}{D-1} P^{+}{}_{\mu_{1}\nu_{1}} P^{+}{}_{\mu_{2}\nu_{2}}\big)\,,
  \\
  & \sum_{\overline{E}} \overline{E}_{\mu_{1}\nu_{1}} \overline{E}_{\mu_{2}\nu_{2}} =a\big( P^{-}{}_{\mu_{1}\mu_{2}} P^{-}{}_{\nu_{1}\nu_{2}} + P^{-}{}_{\mu_{1}\nu_{2}} P^{-}{}_{\nu_{1}\mu_{2}} - \frac{2}{D-1} P^{-}{}_{\mu_{1}\nu_{1}} P^{-}{}_{\mu_{2}\nu_{2}}\big),
\end{aligned}\label{}
\end{equation}
where 
\begin{equation}
\begin{aligned}
  P^{+}{}_{\mu\nu} &= \eta_{\mu\nu} + \frac{\alpha'}{4} k_{\mu} k_{\nu}
  \\
  P^{-}{}_{\mu\nu} &= \eta_{\mu\nu} - \frac{\alpha'}{4} k_{\mu} k_{\nu}
\end{aligned}\label{}
\end{equation}
and the normalization  constant $a$ is fixed by consistency.

We now read off the residue of each amplitude in (\ref{3pt^2}) and compare with the residue of the four-dilaton amplitude in (\ref{4pointdilaton}) for all  $s$-channel poles.  The four-dilaton amplitude constructed from the square of three-point amplitudes reduces to 

\begin{equation}
\begin{aligned}
  	\mbox{Res}_{s=0} \, M_{0} & = -g_c^6 C^2_{S^2}\frac{t^{2}}{16}
  	\\
  	\mbox{Res}_{s=+\frac{4}{\alpha'}} M_{\frac{4}{\alpha'}} & = g_c^6 C^2_{S^2} \frac{(8 + 5 \alpha't) (12 + 5 \alpha't)}{288}a
  	\\
  	\mbox{Res}_{s=-\frac{4}{\alpha'}} M_{-\frac{4}{\alpha'}} & = g_c^6 C^2_{S^2} \frac{(-8 + 5\alpha't) (-12 + 5 \alpha' t)}{ 288}a\,.
\end{aligned}\label{3p2}
\end{equation}
This result is consistent with the direct  computation (\ref{4pointdilaton})
\begin{equation}
\begin{aligned}
  \mbox{Res}_{s=0} M^{\scriptscriptstyle D}_4 &= -g_c^4 C_{S^2}\pi \frac{t^{2}}{2\alpha'}
  	\\
  \mbox{Res}_{s=+ \frac{4}{\alpha'}} M^{\scriptscriptstyle D}_4 &=g_c^4 C_{S^2}\pi \frac{(8 + 5 \alpha't) (12 + 5 \alpha't)}{144 \alpha'}
  	\\
  \mbox{Res}_{s=-\frac{4}{\alpha'}} M^{\scriptscriptstyle D}_4 &=g_c^4 C_{S^2}\pi\frac{(-8 + 5\alpha't) (-12 + 5 \alpha' t)}{144\alpha'}\, .
\end{aligned}\label{4p}
\end{equation}
By comparing (\ref{3p2}) and (\ref{4p})  we find 
\be
C_{S^2}=\frac{8\pi}{\alpha' g_c^2}\ \qquad \text{and} \qquad a=\frac{1}{4} .
\ee
 It is worth to remark that both results match  only at the  critical dimension, $D=26$, and when the condition (\ref{separateequ}) holds. Similar computation was presented in \cite{Leite:2016fno} but after assuming (\ref{C1}) as a quantum consistency condition. Notice, however, that here we find  the critical dimension and the condition (\ref{separateequ}) by applying the standard canonical quantization of string theory for the new vacuum. The validity of (\ref{C1}) is already ensured by the quantum consistency of the quantization scheme.


\section{Four-point scattering amplitudes}\label{4pointappendix}
In this section, we show the detailed form of the $F_{L}{}^{\mu_{1}\mu_{2}\mu_{3}\mu_{4}}$ and $F_{R}{}^{\nu_{1}\nu_{2}\nu_{3}\nu_{4}}$ in (\ref{schem_form}). 
The four-points function of dilatons can be obtained from (\ref{fullA})
\begin{equation}
\begin{aligned}
	\big\langle V_{D}(1) &V_{D}(2) V_{D}(3) V_{D}(4) \big\rangle 
	\\
	&= \varepsilon^{\scriptscriptstyle (D)}_{1 \mu_1\nu_1} \cdots \varepsilon^{\scriptscriptstyle (D)}_{4 \mu_4\nu_4} \frac{\partial}{\partial \xi_{1\mu_1}} \frac{\partial}{\partial \overline{\xi}_{1\nu_1}} \cdots \frac{\partial}{\partial \xi_{4\mu_4}} \frac{\partial}{\partial \overline{\xi}_{4\mu_4}}\cA_{M}(k_i, z_i, \overline{z}_i,\xi_i,\overline{\xi}_i) |_{\xi_i = \overline{\xi}_i=0}\,.
\end{aligned}
\label{}\end{equation}
This can be separated into left-moving and right-moving sectors
\begin{displaymath}
  A_{4} = \varepsilon^{\scriptscriptstyle (D)}_{1 \mu_1\nu_1} \cdots \varepsilon^{\scriptscriptstyle (D)}_{4 \mu_4\nu_4} F_{R}{}^{\mu_1 \mu_2 \mu_3 \mu_4}(z_1,z_2,z_3,z_4) F_{L}{}^{\nu_1 \nu_2 \nu_3 \nu_4}(\overline{z}_1,\overline{z}_2,\overline{z}_3,\overline{z}_4)
\end{displaymath}
where 
\begin{equation}
  	F_{R}{}^{\mu_1 \mu_2 \mu_3 \mu_4} = \frac{\partial}{\partial \xi_{1\mu_1}} \frac{\partial}{\partial \xi_{2\mu_2}} \frac{\partial}{\partial \xi_{3\mu_3}} \frac{\partial}{\partial \xi_{4\mu_4}} \left\langle T_{R}\big[\cV_{R}(1) \cV_{R}(2) \cV_{R}(3) \cV_{R}(4)\big]\right\rangle|_{\xi_i=0}\,,
\end{equation}
and 
\begin{equation}
    F_{L}{}^{\nu_1 \nu_2 \nu_3 \nu_4} = \frac{\partial}{\partial \overline{\xi}_{1\nu_1}} \frac{\partial}{\partial \overline{\xi}_{2\nu_2}} \frac{\partial}{\partial \overline{\xi}_{3\nu_3}} \frac{\partial}{\partial \overline{\xi}_{4\nu_4}} \left\langle T_{L} \big[\cV_{L}(1) \cV_{L}(2) \cV_{L}(3) \cV_{L}(4)\big]\right\rangle|_{\overline{\xi}_i=0}\,.
\label{}\end{equation}
The $T_{R}$ and $T_{L}$ are right-moving and left-moving time-ordering. If we substitute the (\ref{rightcontri}) and (\ref{leftcontri}), then we have the right-moving sector
\begin{equation}
\begin{aligned}
  F_{R}&{}^{\mu_1 \mu_2 \mu_3 \mu_4}(z_1,z_2,z_3,z_4) 
  \\
  &= \big(\frac{\alpha'}{2}\big)^2 \prod_{i<j}(z_{ij})^{\frac{\alpha'}{2}k_i\cdot k_j}\Big[~ \frac{\eta_{\mu_1\mu_4}\eta_{\mu_2\mu_3}}{z_{14}^2 z_{23}^2}  + \frac{\eta_{\mu_1 \mu_3} \eta_{\mu_2 \mu_4}}{z_{13}^{2} z_{24}^{2}} +  \frac{\eta_{\mu_1 \mu_2} \eta_{\mu_3 \mu_4}}{z_{12}^{2} z_{34}^{2}}
\\
&\qquad + \frac{\alpha'}{2} \Big( -\frac{\eta_{\mu_2 \mu_3}}{z_{23}^{2}} \big( \frac{k_{1 \mu_4}}{z_{14}} +\frac{k_{2 \mu_4}}{z_{24}} +\frac{k_{3 \mu_4}}{z_{34}}\big) \big( \frac{k_{2 \mu_1}}{z_{12}} + \frac{k_{3 \mu_1}}{z_{13}} + \frac{k_{4 \mu_1}}{z_{14}} \big)
\\
& \qquad\qquad\quad - \frac{\eta_{\mu_1 \mu_3}}{z_{13}^2} \big( \frac{k_{1\mu_4}}{z_{14}} +\frac{k_{2\mu_4}}{z_{24}} +\frac{k_{3\mu_4}}{z_{34}} \big) \big( - \frac{k_{1\mu_2}}{z_{12}} +\frac{k_{3\mu_2}}{z_{23}} +\frac{k_{4\mu_2}}{z_{24}} \big)
\\
&\qquad\qquad\quad + \frac{\eta_{\mu_3\mu_4}}{z_{34}^2} \big( \frac{k_{2\mu_1}}{z_{12}} +\frac{k_{3\mu_1}}{z_{13}} +\frac{k_{4\mu_1}}{z_{14}}\big) \big( -\frac{k_{1\mu_2}}{z_{12}} +\frac{k_{3\mu_2}}{z_{23}}+\frac{k_{4\mu_2}}{z_{24}} \big)
\\
&\qquad\qquad\quad + \frac{\eta_{\mu_1\mu_2}}{z_{12}^2} \big(-\frac{k_{1\mu_4}}{z_{14}} -\frac{k_{2\mu_4}}{z_{24}} -\frac{k_{3\mu_4}}{z_{34}} \big) \big( -\frac{k_{1\mu_3}}{z_{13}} -\frac{k_{2\mu_3}}{z_{23}} +\frac{k_{4\mu_3}}{z_{34}} \big)
\\
&\qquad\qquad\quad + \frac{\eta_{\mu_2\mu_4}}{z_{24}^2} \big( \frac{k_{2\mu_1}}{z_{12}} +\frac{k_{3\mu_1}}{z_{13}} +\frac{k_{4\mu_1}}{z_{14}} \big) \big( -\frac{k_{1\mu_3}}{z_{13}} -\frac{k_{2\mu_3}}{z_{23}} +\frac{k_{4\mu_3}}{z_{34}} \big)
\\
&\qquad\qquad\quad + \frac{\eta_{\mu_1\mu_4}}{z_{14}^2} \big( -\frac{k_{1\mu_2}}{z_{12}} +\frac{k_{3\mu_2}}{z_{23}} +\frac{k_{4\mu_2}}{z_{24}} \big) \big( -\frac{k_{1\mu_3}}{z_{13}} -\frac{k_{2\mu_3}}{z_{23}} +\frac{k_{4\mu_3}}{z_{34}} \big)\Big)
\\
&\qquad + \big(\frac{\alpha'}{2}\big)^2 \big(-\frac{k_{1\mu_4}}{z_{14}} -\frac{k_{2\mu_4}}{z_{24}} -\frac{k_{3\mu_4}}{z_{34}}\big) \big( \frac{k_{2\mu_1}}{z_{12}} +\frac{k_{3\mu_1}}{z_{13}} +\frac{k_{4\mu_1}}{z_{14}} \big) 
\\ 
& \qquad \quad \qquad ~\times  ( -\frac{k_{1\mu_2}}{z_{12}} +\frac{k_{3\mu_2}}{z_{23}} +\frac{k_{4\mu_2}}{z_{24}} \big)  \big( -\frac{k_{1\mu_3}}{z_{13}} -\frac{k_{2\mu_3}}{z_{23}} +\frac{k_{4\mu_3}}{z_{34}}\big)\Big] \,,
\end{aligned}\label{}
\end{equation}
where $z_{ij} = z_i - z_j$, and left-moving sector 
\begin{equation}
\begin{aligned}
  F_{L}&{}^{\mu_1 \mu_2 \mu_3 \mu_4}(\overline{z}_1,\overline{z}_2,\overline{z}_3,\overline{z}_4) 
  \\
  &= \big(\frac{\alpha'}{2}\big)^2 \prod_{i<j}(\overline{z}_{ij})^{-\frac{\alpha'}{2}k_i\cdot k_j}\Big[~ \frac{\eta_{\mu_1\mu_4}\eta_{\mu_2\mu_3}}{\overline{z}_{14}^2 \overline{z}_{23}^2}  + \frac{\eta_{\mu_1 \mu_3} \eta_{\mu_2 \mu_4}}{\overline{z}_{13}^{2} \overline{z}_{24}^{2}} +  \frac{\eta_{\mu_1 \mu_2} \eta_{\mu_3 \mu_4}}{\overline{z}_{12}^{2} \overline{z}_{34}^{2}}
\\
&\qquad + \frac{\alpha'}{2} \Big( \, \, \frac{\eta_{\mu_2 \mu_3}}{\overline{z}_{23}^{2}} \big( \frac{k_{1 \mu_4}}{\overline{z}_{14}} +\frac{k_{2 \mu_4}}{\overline{z}_{24}} +\frac{k_{3 \mu_4}}{\overline{z}_{34}}\big) \big( \frac{k_{2 \mu_1}}{\overline{z}_{12}} + \frac{k_{3 \mu_1}}{\overline{z}_{13}} + \frac{k_{4 \mu_1}}{\overline{z}_{14}} \big)
\\
& \qquad\quad\quad + \frac{\eta_{\mu_1 \mu_3}}{\overline{z}_{13}^2} \big( \frac{k_{1\mu_4}}{\overline{z}_{14}} +\frac{k_{2\mu_4}}{\overline{z}_{24}} +\frac{k_{3\mu_4}}{\overline{z}_{34}} \big) \big( - \frac{k_{1\mu_2}}{\overline{z}_{12}} +\frac{k_{3\mu_2}}{\overline{z}_{23}} +\frac{k_{4\mu_2}}{\overline{z}_{24}} \big)
\\
&\qquad\quad\quad -\frac{\eta_{\mu_3\mu_4}}{\overline{z}_{34}^2} \big( \frac{k_{2\mu_1}}{\overline{z}_{12}} +\frac{k_{3\mu_1}}{\overline{z}_{13}} +\frac{k_{4\mu_1}}{\overline{z}_{14}}\big) \big( -\frac{k_{1\mu_2}}{\overline{z}_{12}} +\frac{k_{3\mu_2}}{\overline{z}_{23}}+\frac{k_{4\mu_2}}{\overline{z}_{24}} \big)
\\
&\qquad\quad\quad - \frac{\eta_{\mu_1\mu_2}}{\overline{z}_{12}^2} \big(-\frac{k_{1\mu_4}}{\overline{z}_{14}} -\frac{k_{2\mu_4}}{\overline{z}_{24}} -\frac{k_{3\mu_4}}{\overline{z}_{34}} \big) \big( -\frac{k_{1\mu_3}}{\overline{z}_{13}} -\frac{k_{2\mu_3}}{\overline{z}_{23}} +\frac{k_{4\mu_3}}{\overline{z}_{34}} \big)
\\
&\qquad\quad\quad - \frac{\eta_{\mu_2\mu_4}}{\overline{z}_{24}^2} \big( \frac{k_{2\mu_1}}{\overline{z}_{12}} +\frac{k_{3\mu_1}}{\overline{z}_{13}} +\frac{k_{4\mu_1}}{\overline{z}_{14}} \big) \big( -\frac{k_{1\mu_3}}{\overline{z}_{13}} -\frac{k_{2\mu_3}}{\overline{z}_{23}} +\frac{k_{4\mu_3}}{\overline{z}_{34}} \big)
\\
&\qquad\quad\quad - \frac{\eta_{\mu_1\mu_4}}{\overline{z}_{14}^2} \big( -\frac{k_{1\mu_2}}{\overline{z}_{12}} +\frac{k_{3\mu_2}}{\overline{z}_{23}} +\frac{k_{4\mu_2}}{\overline{z}_{24}} \big) \big( -\frac{k_{1\mu_3}}{\overline{z}_{13}} -\frac{k_{2\mu_3}}{\overline{z}_{23}} +\frac{k_{4\mu_3}}{\overline{z}_{34}} \big)\Big)
\\
&\qquad + \Big(\frac{\alpha'}{2}\Big)^2 \big(\frac{k_{1\mu_4}}{\overline{z}_{14}} +\frac{k_{2\mu_4}}{\overline{z}_{24}} +\frac{k_{3\mu_4}}{\overline{z}_{34}}\big) \big( \frac{k_{2\mu_1}}{\overline{z}_{12}} +\frac{k_{3\mu_1}}{\overline{z}_{13}} +\frac{k_{4\mu_1}}{\overline{z}_{14}} \big) 
\\ 
& \quad \qquad \qquad ~\times \big( -\frac{k_{1\mu_2}}{\overline{z}_{12}} +\frac{k_{3\mu_2}}{\overline{z}_{23}} +\frac{k_{4\mu_2}}{\overline{z}_{24}} \big)  \big( \frac{k_{1\mu_3}}{\overline{z}_{13}} +\frac{k_{2\mu_3}}{\overline{z}_{23}} -\frac{k_{4\mu_3}}{\overline{z}_{34}}\big)\Big] \,,
\end{aligned}\label{}
\end{equation}
\noindent Under the $z_1\to \infty$ limit and setting $z_2=1$, $z_3=z$ and $z_4=0$ , one can show that the surviving terms  can be rearranged as
\begin{equation}
\begin{aligned}
  K_{R}(z;k_{i}) = \,& P_{1}(k_{i}) +\frac{P_{2}(k_{i})}{z^2}+ \frac{P_{3}(k_{i})}{(1-z)^2} + \frac{Q_{1}(k_{i})}{z} + \frac{Q_{2}(k_{i})}{1-z}  + \frac{Q_{3}(k_{i})}{z(1-z)}\\
&  + \frac{Q_{4}(k_{i})}{z(1-z)^2} + \frac{Q_{5}(k_{i})}{z^2(1-z)} + \frac{z\ Q_{6}(k_{i})}{(1-z)}+ \frac{z\ Q_{7}(k_{i})}{(1-z)^2}   \,,
\end{aligned}\label{Kine1}
\end{equation}
and 
\begin{equation}
\begin{aligned}
  K_{L}(\overline{z};k_{i}) = \,& \overline{P}_{1}(k_{i}) +\frac{\overline{P}_{2}(k_{i})}{\overline{z}^2}+ \frac{\overline{P}_{3}(k_{i})}{(1-\overline{z})^2} + \frac{\overline{Q}_{1}(k_{i})}{\overline{z}} + \frac{\overline{Q}_{2}(k_{i})}{1-\overline{z}}  + \frac{\overline{Q}_{3}(k_{i})}{\overline{z}(1-\overline{z})}\\
&  + \frac{\overline{Q}_{4}(k_{i})}{\overline{z}(1-\overline{z})^2} + \frac{\overline{Q}_{5}(k_{i})}{\overline{z}^2(1-\overline{z})} + \frac{\overline{z}\ \overline{Q}_{6}(k_{i})}{(1-\overline{z})}+ \frac{\overline{z}\ \overline{Q}_{7}(k_{i})}{(1-\overline{z})^2} \,.
\end{aligned}\label{Kine2}
\end{equation}
Here,
\begin{equation}
\begin{aligned}
  P_{1}{}^{\mu_{1}\mu_{2}\mu_{3}\mu_{4}} &= \eta^{\mu_1 \mu_3} \eta^{\mu_2 \mu_4} + \frac{\alpha'}{2}\big(- \eta^{\mu_1 \mu_3}  k_{2}^{\mu_4} k_{4}^{\mu_2}+\eta^{\mu_2\mu_4}k_3^{\mu_1}k_4^{\mu_3} \big)- \big(\frac{\alpha'}{2}\big)^2 k_2^{\mu_4}k_3^{\mu_1}k_4^{\mu_2}k_4^{\mu_3}\,,
  \\
  P_{2}{}^{\mu_{1}\mu_{2}\mu_{3}\mu_{4}} &= \eta^{\mu_1 \mu_2} \eta^{\mu_3 \mu_4} +\frac{\alpha'}{2}\big(- \eta^{\mu_1 \mu_2} k_3^{\mu_4} k_4^{\mu_3}+\eta^{\mu_3\mu_4 }  k_{2}^{\mu_1} k_{4}^{\mu_2}\big)- \big(\frac{\alpha'}{2}\big)^2 k_2^{\mu_1}k_3^{\mu_4}k_4^{\mu_2}k_4^{\mu_3}\,,
  \\
  P_3{}^{\mu_{1}\mu_{2}\mu_{3}\mu_{4}} &= \eta^{\mu_1 \mu_4} \eta^{\mu_2 \mu_3} + \frac{\alpha'}{2}\big(- \eta^{\mu_1 \mu_4} k_{2}^{\mu_3} k_{3}^{\mu_2}- \eta^{\mu_2 \mu_3} k_{2}^{\mu_1} k_{2}^{\mu_4}- \eta^{\mu_2 \mu_3} k_{3}^{\mu_1} k_{3}^{\mu_4} \big)\\
  & \qquad \qquad \qquad +\big(\frac{\alpha'}{2}\big)^2 \big( k_2^{\mu_1}k_2^{\mu_3}k_2^{\mu_4}k_3^{\mu_2}+ k_2^{\mu_3}k_3^{\mu_1}k_3^{\mu_2}k_3^{\mu_4} \big)\,,
  \\
  Q_1{}^{\mu_{1}\mu_{2}\mu_{3}\mu_{4}} &= \frac{\alpha'}{2} \big(\eta^{\mu_1 \mu_4} k_4^{\mu_2} k_4^{\mu_3} - \eta^{\mu_1 \mu_3} k_3^{\mu_4} k_4^{\mu_2} - \eta^{\mu_1 \mu_2} k_2^{\mu_4} k_4^{\mu_3}+\eta^{\mu_3 \mu_4} k_3^{\mu_1} k_4^{\mu_2} +\eta^{\mu_2 \mu_4} k_2^{\mu_1} k_4^{\mu_3} \big)\\
  & +\big(\frac{\alpha'}{2}\big)^2 \big( -k_2^{\mu_1}k_2^{\mu_4}k_4^{\mu_2}k_4^{\mu_3}- k_3^{\mu_1}k_3^{\mu_4}k_4^{\mu_2}k_4^{\mu_3} \big)\,,
  \\
  Q_2{}^{\mu_{1}\mu_{2}\mu_{3}\mu_{4}} &= \frac{\alpha'}{2} \big(\eta^{\mu_1 \mu_2} k_{2}^{\mu_4} k_{2}^{\mu_3} - \eta^{\mu_1 \mu_3} k_{2}^{\mu_4} k_{3}^{\mu_2} - \eta^{\mu_1 \mu_4} k_{2}^{\mu_3} k_{4}^{\mu_2} - \eta^{\mu_2 \mu_4} k_{2}^{\mu_1} k_{2}^{\mu_3}  \big)\\
 & +\big(\frac{\alpha'}{2}\big)^2 \big( k_2^{\mu_1}k_2^{\mu_3}k_2^{\mu_4}k_4^{\mu_2}+ k_2^{\mu_3}k_3^{\mu_1}k_3^{\mu_4}k_4^{\mu_2}-k_2^{\mu_4}k_3^{\mu_1}k_3^{\mu_2}k_4^{\mu_3} \big)\,,
  \\
  Q_3{}^{\mu_{1}\mu_{2}\mu_{3}\mu_{4}} &= \frac{\alpha'}{2} \big( \eta^{\mu_1 \mu_2} k_{2}^{\mu_3} k_{3}^{\mu_4} + \eta^{\mu_1 \mu_4} k_{3}^{\mu_2} k_{4}^{\mu_3} - \eta^{\mu_1 \mu_3} k_{3}^{\mu_2} k_{3}^{\mu_4}+\eta^{\mu_3 \mu_4} k_{3}^{\mu_1} k_{3}^{\mu_2}\big)\\
  & +\big(\frac{\alpha'}{2}\big)^2 \big( k_2^{\mu_1}k_2^{\mu_3}k_3^{\mu_4}k_4^{\mu_2}- k_2^{\mu_1}k_2^{\mu_4}k_3^{\mu_2}k_4^{\mu_3}-k_3^{\mu_1}k_3^{\mu_2}k_3^{\mu_4}k_4^{\mu_3} \big)\,,
  \\
  Q_4{}^{\mu_{1}\mu_{2}\mu_{3}\mu_{4}} & = - \frac{\alpha'}{2} \eta^{\mu_2\mu_3} k_2^{\mu_1} k_3^{\mu_4} +\big( \frac{\alpha'}{2}\big)^2 k_2^{\mu_1} k_2^{\mu_3} k_3^{\mu_2} k_3^{\mu_4}\,,
   \\
  Q_5{}^{\mu_{1}\mu_{2}\mu_{3}\mu_{4}} & = + \frac{\alpha'}{2} \eta^{\mu_3\mu_4} k_2^{\mu_1} k_3^{\mu_2} -\big( \frac{\alpha'}{2}\big)^2 k_2^{\mu_1} k_3^{\mu_2} k_3^{\mu_4} k_4^{\mu_3}\,,
   \\
  Q_6{}^{\mu_{1}\mu_{2}\mu_{3}\mu_{4}} & = - \frac{\alpha'}{2} \eta^{\mu_2\mu_4} k_2^{\mu_3} k_3^{\mu_1} +\big( \frac{\alpha'}{2}\big)^2 k_2^{\mu_3} k_2^{\mu_4} k_3^{\mu_1} k_4^{\mu_2}\,,
   \\
  Q_7{}^{\mu_{1}\mu_{2}\mu_{3}\mu_{4}} & = - \frac{\alpha'}{2} \eta^{\mu_2\mu_3} k_2^{\mu_4} k_3^{\mu_1} +\big( \frac{\alpha'}{2}\big)^2 k_2^{\mu_3} k_2^{\mu_4} k_3^{\mu_1} k_3^{\mu_2}\,.
\end{aligned}\label{PQ}
\end{equation}
We can get $\overline{P}_{i}$ and $\overline{Q}_{i}$ by substituting $\alpha'\to -\alpha'$ from the above definitions. 


\end{document}